\documentclass[nofootinbib,prd,a4paper,showpacs]{revtex4} 
\usepackage{epsfig}

\def\nn{\nonumber}        

\newcommand{\bm}[1]{\mbox{\boldmath $#1$}}

\newcommand{\open}{{<\kern -0.3 em{\scriptscriptstyle )}}}

\newcommand{\nslash}{\kern 0.2 em n\kern -0.45em /}
\newcommand{\Pslash}{\kern 0.2 em P\kern -0.56em \raisebox{0.3ex}{/}}
\newcommand{\pslash}{\kern 0.2 em p\kern -0.4em /}
\newcommand{\kslash}{\kern 0.2 em k\kern -0.45em /}
\newcommand{\Sslash}{\kern 0.2 em S\kern -0.56em \raisebox{0.3ex}{/}}
\newcommand{\slsh}[1]{\mbox{$\not\! #1$}}

\newcommand{\eq}{\begin{equation}}
\newcommand{\ee}{\end{equation}}
\newcommand{\beq}{\begin{equation}}
\newcommand{\eeq}{\end{equation}}
\newcommand{\ba}{\begin{eqnarray}}
\newcommand{\ea}{\end{eqnarray}}
\newcommand{\eqa}{\begin{eqnarray}}
\newcommand{\eea}{\end{eqnarray}}
\newcommand{\ph}{{\rule{0mm}{3mm}}}
\newcommand{\half}{\textstyle{1\over 2}}
\newcommand{\sumint}{\kern 0.2 em {\textstyle\sum} \kern -1.1 em \int}

\newcommand{\g}{\gamma}

\newcommand{\simorder}{\raisebox{-4pt}{$\, \stackrel{\textstyle >}{\sim} \,$}}
\newcommand{\simorderr}{\raisebox{-4pt}{$\, \stackrel{\textstyle <}{\sim} \,$}}
\newcommand{\bkt}{\bm{k}_{T}}
\newcommand{\la}{\langle}
\newcommand{\ra}{\rangle}
\newcommand{\amp}[1]{\la #1 \ra}

\begin{document} 

\title{Angular dependences in inclusive two-hadron production at BELLE}

\author{Dani\"el Boer}
\email{D.Boer@few.vu.nl}
\affiliation{Department of Physics and Astronomy, 
Vrije Universiteit Amsterdam, \\
De Boelelaan 1081, 1081 HV Amsterdam, The Netherlands}

\date{\today}

\begin{abstract}
A collection of results is presented relevant for the analysis of
azimuthal asymmetries in inclusive two-hadron production at BELLE. The aim of
this overview is to provide theoretical ingredients necessary to extract 
the Collins effect fragmentation function. The latter arises within the 
Collins-Soper factorization formalism, which describes both the 
transverse momentum and $Q^2$ dependence of the cross section and its angular
dependences at low and moderate transverse momentum. Since the Collins effect
is not the only source of angular dependences, a discussion of various other
effects is included. This concerns higher twist contributions, 
photon--$Z$-boson interference effects, radiative corrections, beam 
polarization and weak decays. Furthermore, different frames, transverse 
momentum weighting and ratios of asymmetries are discussed. 
These issues are all of relevance for the unambiguous measurement of the 
Collins effect.  
\end{abstract}

\pacs{13.66.Bc,13.87.Fh,13.88.+e} 

\maketitle


\section{Introduction}

The BELLE experiment at KEK in Japan measures with very high luminosity the
process of electron-positron annihilation in collisions of 8.0 GeV electrons
and 3.5 GeV positrons. The center of mass energy is selected to be
on-resonance of the $\Upsilon(4S)$ meson, which has a mass of $10.5800 \pm
0.0035 $ GeV and which decays more than 96\% of the time into $B \bar{B}$
meson pairs. The main aim of the measurements of $B$ and $\bar B$ decays is to
study $CP$ violation. Besides this goal, there are other interesting studies
that can be performed at BELLE\footnote{We have chosen to focus on BELLE, but
  the observables to be discussed can of course also be studied at other $e^+
  e^-$ colliders.}  and for which also the off-resonance data are useful
(which in the case of BELLE are taken $60$ MeV below resonance). The
acquisition of off-resonance data is mostly used for background studies, but
is also of interest for physics studies that are not $b$-quark specific. This
overview discusses such a case, namely the study of azimuthal asymmetries in
the inclusive production of two almost back-to-back hadrons, $e^+e^- \to h_1
\, h_2\, X$. There are several effects that can cause such asymmetries and one
would like to disentangle them in order to isolate perturbative from
nonperturbative effects. In the latter type of effects quark spin is expected
to play a nontrivial role via the so-called Collins effect
\cite{Collins-93b}. By measuring azimuthal asymmetries it may be possible to
obtain trustworthy quantitative knowledge on this type of effect, for instance
from the BELLE data. Recently, the first results from BELLE were published
\cite{Abe:2005zx} and the purpose of this overview is to discuss the
theoretical aspects of such type of study.

The Collins effect was first discussed in the context of semi-inclusive deep
inelastic scattering (SIDIS) of leptons off transversely polarized protons, as
a means to access transversity \cite{Collins-93b}. Transversity
\cite{Ralst-S-79} describes the extent to which quarks are transversely
polarized inside a proton that is polarized transversely to the probing
particle, which in the case of SIDIS is a virtual photon. The Collins effect
describes the angular asymmetry in the distribution of hadrons produced from a
transversely polarized fragmenting quark. Via this effect the transverse
polarization of the struck proton results in an asymmetric distribution of
final state hadrons. If sufficiently large, the Collins effect would thus
allow for a measurement of transversity and subsequently of the tensor
charge, the fundamental charge that can only be measured through transversity.

The first nonzero Collins effect asymmetry in polarized SIDIS has been
observed by the HERMES experiment \cite{Airapetian:2004tw} (using a deuteron 
target the COMPASS experiment obtained a result consistent with zero
\cite{Ageev:2006da}, presumably due to cancellations between proton and
neutron contributions). The HERMES result indicates that both transversity and 
the Collins effect are nonzero. 
For an extraction of transversity from those SIDIS data a
separate measurement of the Collins effect fragmentation function needs to be
performed. This can only be done in the process $e^+e^- \to h_1 \, h_2\, X$
\cite{BJM97} 
and motivates the BELLE efforts concerning the measurement of this process.
Some earlier attempt to use LEP1 data has been undertaken \cite{Efremov}, but
without study of systematic effects and still remains preliminary. Moreover,
as will be discussed, it is likely that the Collins effect asymmetry in
$e^+e^- \to h_1 \, h_2\, X$ has a powerlike fall-off behavior with energy,
which would favor an extraction at BELLE over LEP1. 

We will study various effects that could lead to azimuthal asymmetries in the
process of interest, $e^+e^- \to h_1 \, h_2\, X$, 
which besides the Collins effect, include electroweak
$\gamma$-$Z$ interference effects, beam polarization effects and radiative
corrections. In order to arrive at an unambiguous 
interpretation of the data the magnitude
and scale dependence of the various effects need to be estimated, of course to
the extent to which that is possible from first principles. The effects will
not be treated simultaneously; combinations of effects will only be considered
when the analysis requires it. As a rule we will ignore effects that are
expected to be smaller than a permille, such as $Z$-$Z$ contributions or beam
polarization in combination with $\gamma$-$Z$ interference. From the possible
contributions considered the higher-twist effects are the least known, but
unfortunately not necessarily below the percent level at BELLE. In cases where
no reliable estimate can be given, such as for twist-4 effects (${\cal
  O}(\Lambda^2/Q^2)$), additional observables or checks may need to be
considered to further exclude competing interpretations of the Collins effect
asymmetry. But despite some uncertainties the isolation of the Collins effect
contribution does seem feasible given the possibilities that BELLE offers. 
The purpose of this overview, which contains several new aspects, 
is to assist and facilitate this endeavor. 
The realistic prospect of extracting the Collins effect fragmentation
function and with it transversity, of which a first result has recently been
obtained \cite{Anselmino:2007fs}, makes the study of azimuthal asymmetries 
at BELLE well worth the effort.  

This overview consists of the following sections and subsections:
\[
\begin{array}{ll}
\ref{sec:Gen} & \text{General angular dependence}\\
\ref{sec:Two} & \text{Two-particle inclusive cross section}\\
\ref{sec:Frames}  &  \text{Frames}\\
\ref{sec:SF}  & \text{Structure functions}\\
\ref{sec:LOxs}  & \text{Leading order cross section}\\
\ref{sec:Integrated}  & \text{Integration over transverse photon momentum}\\
\ref{sec:Unintegrated}  & \text{Unintegrated cross section}\\
\ref{sec:Weigh}  & \text{Weighted cross sections}\\
\ref{sec:Estimate}  & \text{Estimate of the $Q_T^2$-weighted Collins effect asymmetry}\\
\ref{sec:Univ}  & \text{Universality of the Collins effect}\\
\ref{sec:HT}  & \text{Higher twist}\\
\ref{sec:EW}  & \text{Electroweak interference effects}\\
\ref{sec:Jet}  & \text{Jet frame asymmetry}\\
\ref{sec:Scale}  & \text{Scale dependence of the Collins effect asymmetry}\\
\ref{sec:CS81}  & \text{Collins-Soper factorization}\\
\ref{sec:Num}  & \text{Numerical study of the $Q^2$ dependence of the Collins
  effect asymmetry}\\
\ref{sec:Tree} & \text{Comparison to tree level}\\
\ref{sec:SNP}  & \text{Nonperturbative Sudakov factor from BELLE}\\
\ref{sec:Rad}  & \text{Radiative corrections}\\
\ref{sec:Weigh2}  & \text{Weighted asymmetry beyond tree level}\\
\ref{sec:Ratios}  & \text{Ratios of asymmetries}\\
\ref{sec:Beampol}  & \text{Beam polarization}\\
\ref{sec:Weak2}  & \text{Weak decays background}\\
\ref{sec:Sum}  & \text{Summary}
\end{array}
\]
\section{\label{sec:Gen}General angular dependence}

We consider $e^- e^+\rightarrow h_1\; h_2 \; X $, where the two leptons (with
momentum $l$ for the $e^-$ and $l^\prime$ for the $e^+$) annihilate into a
photon (or $Z$ boson) with momentum $q = l + l^\prime$. This photon momentum 
sets the scale $Q$, where $Q^2 \equiv q^2$, which is much larger than 
characteristic hadronic scales. 
Denoting the momentum of outgoing hadrons by $P_h$ ($h$ = 1, 2) we use
invariants $z_h$ = $2P_h\cdot q/Q^2$. We will consider the case of unpolarized
leptons and hadrons, although in Sec.\ \ref{sec:Beampol} 
we will turn to the issue of
transverse beam polarization due to the Sokolov-Ternov effect. We will work in
the limit where $Q^2$ and $P_h\cdot q$ are large, keeping the ratios $z_h$
finite. We will consider the case where the two hadronic momenta $P_1$ and
$P_2$ do not belong to the same jet (i.e., $P_1\cdot P_2$ is of order $Q^2$).

There is a considerable literature on angular correlations for three-jet
events in electron-positron annihilation, e.g.\
\cite{Kramer:1978uj,Laermann:1979zk,Koller:1979df,Johnson:1981td,Shakhnazarian:1982jh,Hagiwara-91,Brandenburg:1997pu,Braunschweig:1989qw,Adeva:1991dx,Abreu:1991rc},
but here we are limiting the discussion to two-jet events exclusively.  

In general, the differential cross section for the process $e^+e^- \to h_1 \,
h_2\, X$ can be written as (see for instance \cite{Hagiwara-91})
\ba
\frac{dN}{d\Omega} \equiv \left(\frac{d \sigma}{d z_1 d z_2 d^2 \bm{q}_T} 
\right)^{-1} \frac{d \sigma}{dz_1 dz_2 d\Omega d^2 \bm{q}_T} & = & F_1 
(1+ \cos^2 \theta) + F_2(1- 3\cos^2\theta)
+ F_3 \cos \theta \nn \\
& & + F_4 \sin 2\theta \cos \phi + F_5 \sin^2 \theta \cos 2\phi
+ F_6 \sin \theta \cos \phi \nn \\[3 mm]
& & + F_7 \sin 2\theta \sin \phi + F_8 \sin^2 \theta
\sin 2\phi + F_9 \sin \theta \sin \phi.
\label{gendecomp}
\ea 
The functions $F_i$ depend on the invariants $z_h$ = $2P_h\cdot q/Q^2$ and 
on $\bm{q}_T^2 \equiv Q_T^2$, 
the squared transverse momentum $\bm{q}_T$ of the photon 
with respect to the two hadrons. 
The angles $\phi$ and $\theta$ are given in the lepton-pair center of
mass frame or equivalently the photon center of mass frame. Precise
definitions and explanations will be given below.

If at high $Q^2$ and $Q_T^2$ collinear factorization of the cross section is
considered, then at tree level (zeroth order in $\alpha_s$) only $F_1,
F_3$ will receive nonzero contributions ($F_3$ only from $\gamma$-$Z$
interference), at first order in $\alpha_s$ $F_1, \ldots, F_6$ receive
contributions and at second order all $F_i$ are nonzero.  All this is assuming
no transverse beam polarization is present. As said, the complication of
transverse beam polarization will be considered in Sec.\ \ref{sec:Beampol}.

Below we will study the differential cross section $d \sigma/dz_1 dz_2 d\Omega
d^2 \bm{q}_T$ in much detail, also at lower values of $Q_T^2$ where collinear
factorization is not the appropriate framework. To set
the notation we will first look at the cross section expression in terms of
the hadron tensor.
 
\section{\label{sec:Two}Two-particle inclusive cross section}

The square of the amplitude for $e^- e^+\rightarrow h_1 \; h_2 \; X $
can be split into a purely leptonic and a purely hadronic part, 
\begin{equation}
|{\cal M}|^2 = \frac{e^4}{Q^4} L_{\mu\nu} H^{\mu\nu} ,
\end{equation}
with the helicity-conserving lepton tensor (neglecting the lepton masses) 
given by
\begin{equation} \label{leptten2}
L_{\mu\nu} (l, l^\prime)
=  2 l_\mu l^\prime_\nu
+ 2 l_\nu l^\prime_\mu - Q^2 g_{\mu\nu}.
\end{equation}
For the case of two observed hadrons in the final state, the product of 
hadronic current matrix elements is written as
\begin{equation}
H_{\mu\nu}(P_X;  P_1; P_2)
= \langle 0 |J_\mu (0)|P_X; P_1; P_2 \rangle
\langle P_X; P_1; P_2 |J_\nu (0)| 0 \rangle .
\end{equation}
The cross section for two-particle inclusive $e^+e^-$ annihilation is given by (including a factor 1/2 from averaging over initial
state polarizations)  
\begin{equation}
\frac{P_1^0 \,P_2^0\,\,d\sigma^{(e^+e^-)}}{d^3P_1\,d^3P_2}
=\frac{\alpha^2}{4\,Q^6} L_{\mu\nu}{\cal W}^{\mu\nu},
\label{cross2}
\end{equation}
with
\begin{equation}
{\cal W}_{\mu\nu}( q;  P_1; P_2) =
\int \frac{d^3 P_X}{(2\pi)^3 2P_X^0}
\delta^4 (q-P_X - P_1 - P_2)
H_{\mu\nu}(P_X;  P_1; P_2).
\label{hadrten2}
\end{equation}

\subsection{\label{sec:Frames}Frames}

For the calculation of the hadron tensor it will be convenient to define {\em
  lightlike\/} directions using the hadronic (or jet) momenta. The two
hadronic momenta $P_1$ and $P_2$ can be
parameterized using the dimensionless lightlike vectors $n_+$ and $n_-$
(satisfying $n_+ \cdot n_- = 1$),
\begin{eqnarray}
&& P_1^\mu \equiv \frac{\zeta_1\tilde Q}{\sqrt{2}}\,n_-^\mu
+ \frac{M_1^2}{\zeta_1\tilde Q\sqrt{2}}\,n_+^\mu, \label{P1}\\
&& P_2^\mu \equiv \frac{M_2^2}{\zeta_2\tilde Q\sqrt{2}}\,n_-^\mu + 
\frac{\zeta_2\tilde Q}{\sqrt{2}}\,n_+^\mu, \label{P2}\\
&&q^\mu \equiv \frac{\tilde Q}{\sqrt{2}}\,n_-^\mu
+ \frac{\tilde Q}{\sqrt{2}}\,n_+^\mu + q_T^\mu, \label{q}
\end{eqnarray}
where $\tilde Q^2$ = $Q^2 + Q_T^2$ with $q_T^2 \equiv -Q_T^2$. When $Q_T^2 \ll
Q^2$ one has $\tilde Q = Q$, $\zeta_1 = z_1$ and $\zeta_2 = z_2$ up to
$Q_T^2/Q^2$ corrections. 

Vectors transverse to $n_+$ and $n_-$ one obtains using the tensors
\begin{eqnarray}
&&g^{\mu\nu}_T \ \equiv \ g^{\mu\nu}
- n_+^{\,\{\mu} n_-^{\nu\}}, \label{gTtensor}\\
&&\epsilon^{\mu\nu}_T \ \equiv
\ \epsilon^{\mu\nu\rho\sigma} n_{+\rho}n_{-\sigma}. \label{epsTtensor}
\end{eqnarray}

The experimental analysis of the azimuthal asymmetries will usually not be
performed in the frame in which the two hadrons are collinear, for which the
above (Sudakov) decomposition into lightlike vectors and transverse parts is
most suited. Instead it is much more common to consider angles in the 
lepton-pair center of mass frame or equivalently, the photon rest frame. 
In this case there is still freedom to select which momentum (or linear
combination of momenta) is used to define the $\hat{z}$ axis, or equivalently,
what determines the perpendicular plane in which the azimuthal angles lie. 
A few choices are common, such as the Gottfried-Jackson frame and the
Collins-Soper frame. 

For the most part of this overview the frame to be employed will be the
$e^+e^-$-annihilation analogue of the so-called Gottfried-Jackson frame
\cite{Gottfried:1964nx}. We will first give the details of this frame. Later
on we will also consider the analogue of the Collins-Soper frame \cite{CS-77}
and a frame where the jet or thrust axis is used to fix the basis (explained
separately in Sec.\ \ref{sec:Jet}). We will alternate between these different
basis sets depending on what is most convenient for the analysis.

\begin{figure}[htb]
\begin{center}
\leavevmode \epsfxsize=10cm \epsfbox{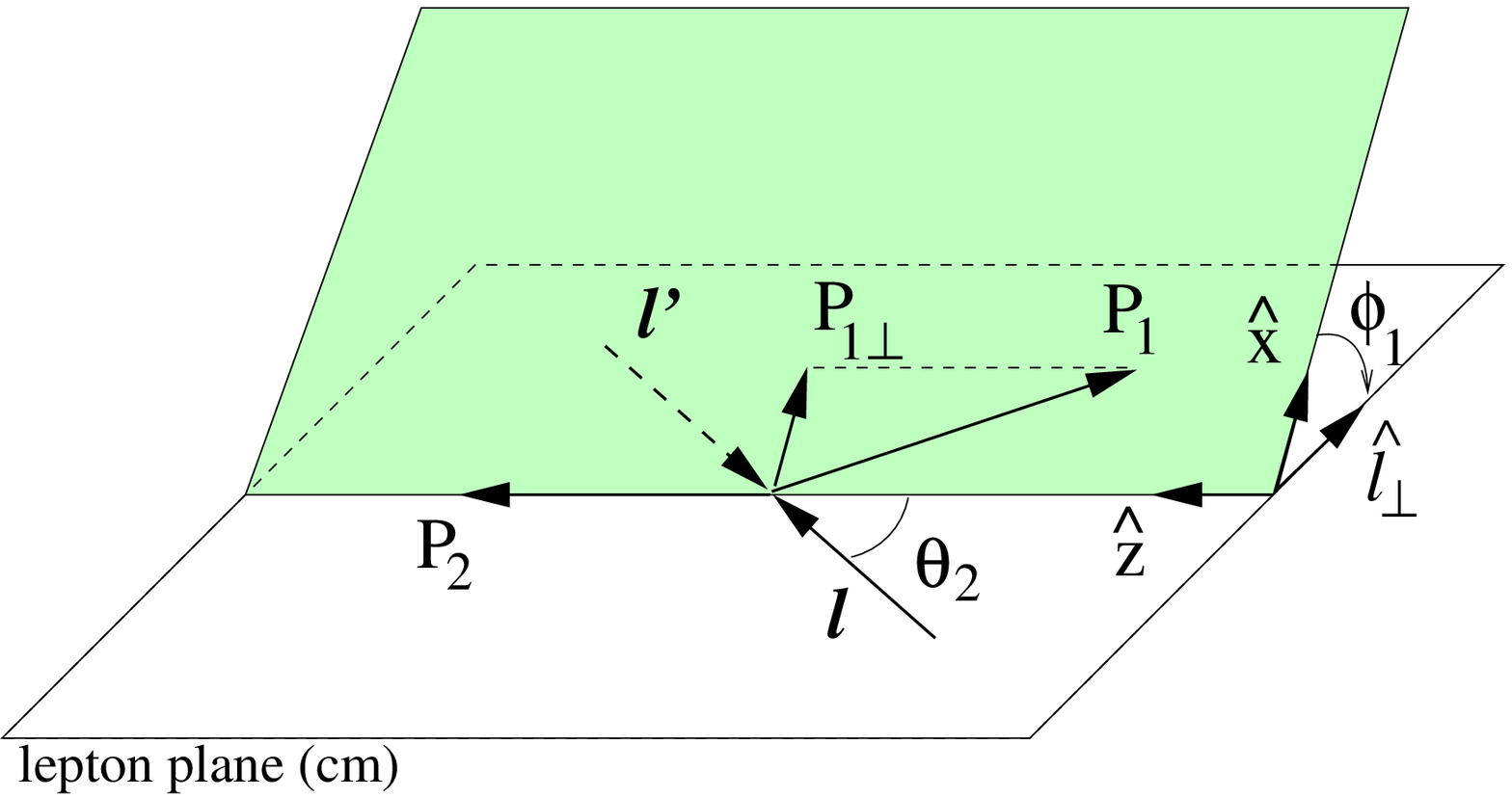}
\vspace{0.2 cm}
\caption{\label{fig:kinannGJ} Kinematics of the annihilation process in the
  lepton center of mass frame (the analogue of the Gottfried-Jackson frame)
  for a back-to-back jet situation. $P_2$ is the momentum of a hadron in one
  jet, $P_1$ is the momentum of a hadron belonging to the other jet.}
\vspace{-5 mm}
\end{center}
\end{figure}

In order to expand the hadron tensor in terms of independent Lorentz
structures (parameterized by structure functions), it is convenient to work
with vectors orthogonal to $q$. A normalized timelike vector is defined by $q$
and a normalized spacelike vector is defined by $\tilde P_i^\mu$ = $P_i^\mu -
(P_i\cdot q/q^2)\,q^\mu$ for one of the outgoing momenta, say $P_2$,
\begin{eqnarray}
\hat t^\mu& \equiv & \frac{q^\mu}{Q},\\
\hat z^\mu & \equiv &
\frac{Q}{P_2\cdot q}\,\tilde P^\mu_2
\ =\   2\,\frac{P_2^\mu}{z_2 Q} - \frac{q^\mu}{Q}. 
\end{eqnarray}
This choice of frame is the analogue of the Gottfried-Jackson frame often
employed for the Drell-Yan process. 
This means that in the lepton-pair center
of mass frame, hadron 2 is moving along the $\hat{\bm{z}}$ direction (see Fig.\
\ref{fig:kinannGJ}). In general, in this frame hadron 1 will have momentum
components orthogonal to $\hat{\bm{z}}$ and $\hat{\bm{t}}$. 

Vectors orthogonal to $\hat z$ and $\hat t$ are obtained with help of the
tensors
\begin{eqnarray}
&& g_{\perp}^{\mu\nu}
\equiv  g^{\mu\nu} -\hat t^\mu \hat t^\nu +
\hat z^\mu \hat z^\nu, \\
&& \epsilon_\perp^{\mu \nu} \equiv
-\epsilon^{\mu \nu \rho \sigma} \hat t_\rho \hat z_\sigma
\ =\ \frac{1}{(P_2\cdot q)}\,\epsilon^{\mu \nu \rho
\sigma} P_{2\,\rho}q_\sigma.
\end{eqnarray}
Since we have chosen hadron 2 to define the longitudinal direction, the
momentum $P_1$ of hadron 1 can be used to express the directions orthogonal to
$\hat t$ and $\hat z$. One obtains $P_{1\perp}^\mu$ =
$g_\perp^{\mu\nu}\,P_{1\nu}$ . We define the normalized vector $\hat h^\mu$ =
$P_{1\perp}^\mu/\vert \bm P_{1\perp} \vert$ and the second orthogonal
direction is given by $\epsilon_\perp^{\mu \nu} \hat h_\nu$. 

Note that the {\em transverse} tensors in Eqs.\ (\ref{gTtensor}) and
(\ref{epsTtensor}) are not identical to the {\em perpendicular} ones defined
above if the transverse momentum of the outgoing hadron 1 does not vanish. The
lightlike directions can easily be expressed in $\hat t$, $\hat z$
and $\hat h$, namely, up to corrections of order $Q_T^2/Q^2$, 
\begin{eqnarray}
n_+^\mu & = & \frac{1}{\sqrt{2}} \left[ \hat t^\mu + \hat z^\mu \right],
\\
n_-^\mu & = & 
\frac{1}{\sqrt{2}} \left[ \hat t^\mu - \hat z^\mu
+ 2\,\frac{Q_T^{}}{Q}\,\hat h^\mu \right]. 
\end{eqnarray}
This shows that the differences between $g^{\mu\nu}_\perp$ and $g^{\mu\nu}_T$
are of order $1/Q$. Especially for the
treatment of azimuthal asymmetries subleading in $1/Q$ (Sec.\ \ref{sec:HT}),
it is important to keep track of these differences. 
We will see however that taking transverse momentum into
account does not automatically lead to suppression. 

Thus far we considered two sets of basis vectors, the first set constructed
from the two hadron momenta ($P_1$ and $P_2$), the second set from the photon
momentum ($q$) and one of the hadron momenta ($P_2$). The respective frames
where the momenta $P_1$ and $P_2$, or $q$ and $P_2$, are collinear are the
natural ones connected to these two sets. One can go from one frame to the
other via a Lorentz transformation that leaves the minus components unchanged
\cite{Levelt-Mulders-94}. In the first frame $q$ has a transverse component
$q_T^{}$, in the second $P_1$ has a perpendicular component $P_{1\perp}$. We
will therefore sometimes refer to them as the ``transverse basis'' and the
``perpendicular basis'', respectively. 
Up to corrections of order $Q_T^2/Q^2$ 
$q_T^\mu$ and $P_{1\perp}^\mu$ are related as follows:
\begin{equation}
P_{1\perp}^\mu = - z_1\,q_T^\mu = z_1\, Q_T\, \hat h^\mu.
\end{equation}

Azimuthal angles will lie inside the plane orthogonal to $\hat{\bm{t}}$ and
$\hat{\bm{z}}$. In particular, $\phi^\ell$ gives the orientation of
$\hat{\bm{l}}_{\perp}$, where $\hat{l}_\perp^\mu$
denotes the normalized perpendicular part of the lepton momentum $l^\mu$. 
The angle
$\phi_1$ is between $\hat{\bm{h}} \propto \bm{P}_{1\perp}$ and 
$\hat{\bm{l}}_\perp$. More specifically, 
\begin{eqnarray}
\hat{l}_\perp \cdot a_\perp &=& - |{\bm a}_\perp | \cos \phi_a,
\label{anglecos} \\
\epsilon^{\mu\nu}_\perp \hat{l}_{\perp\mu} a_{\perp\nu}&=& |{\bm a}_\perp | 
\sin 
\phi_a,
\label{anglesin}
\end{eqnarray}
for a generic vector $a$. The convention for the epsilon tensor used is
$\epsilon^{0123} = 1$. 

Sometimes it may be convenient to choose a different (rotated) set of basis
vectors in the lepton center of mass frame, the Collins-Soper frame. 
For comparison let us denote the
basis vectors of the Gottfried-Jackson (GJ) 
frame for $e^+ e^- \to h_1 h_2 X$ as:
\ba
\hat{t}_{GJ}^\mu & = & \frac{q^\mu}{Q},\\
\hat{z}_{GJ}^\mu & = & \frac{Q}{P_2\cdot q} \tilde{P}_2^\mu,\\
\hat{x}_{GJ}^\mu & = & \frac{2Q}{\tilde{s} Q_T} \left((P_1\cdot \hat{z}_{GJ}) \tilde{P}_2^\mu - (P_2\cdot \hat{z}_{GJ})
  \tilde{P}_1^\mu \right), 
\ea 
where $\tilde{s}= (P_1+P_2)^2$. Here $\hat{x}_{GJ}$ corresponds to $\hat{h}$
and $\hat{y}_{GJ}$ is defined implicitly by requiring a right-handed basis.

The basis for the Collins-Soper (CS) frame for 
$e^+ e^- \to h_1 h_2 X$ is defined as
\cite{Soper:1982wc}:
\ba
\hat{t}_{CS}^\mu & = & \frac{q^\mu}{Q},\\
\hat{z}_{CS}^\mu & = & \frac{2}{\tilde{s} \tilde{Q}} \left( 
(P_1\cdot q) \tilde{P}_2^\mu - (P_2\cdot q) \tilde{P}_1^\mu \right),\\
\hat{x}_{CS}^\mu & = & \frac{2Q}{\tilde{s} Q_T \tilde{Q}} \left(
(P_1\cdot q) \tilde{P}_2^\mu + (P_2\cdot q) \tilde{P}_1^\mu \right).
\ea
Note that we are keeping terms of order $Q_T^2/Q^2$, but not order $M_i^2/Q^2$. 
Throughout this paper we will neglect target mass corrections. 

\begin{figure}[htb]
\begin{center}
\leavevmode \epsfxsize=10cm \epsfbox{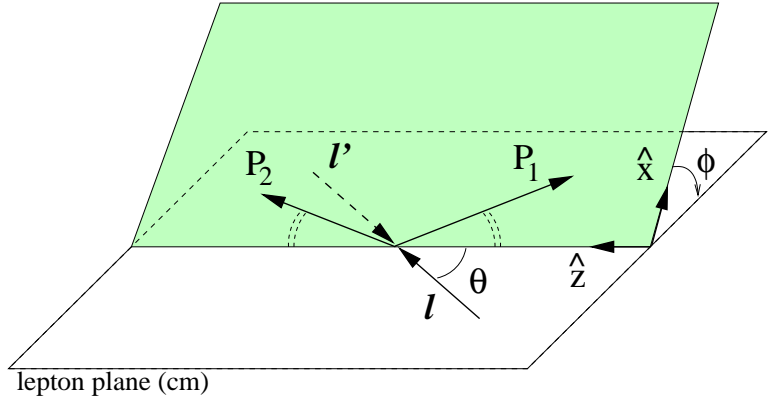}
\vspace{0.2 cm}
\caption{\label{fig:kinannCS} Kinematics of the annihilation process in the
  lepton center of mass frame, the analogue of the Collins-Soper frame.}
\vspace{-5 mm}
\end{center}
\end{figure}

In the lepton-pair center of mass frame the $\hat{\bm{z}}$ axis now points in
the direction that bisects the three-vectors $\bm{P}_2$ and $-\bm{P}_1$ 
(see Fig.\ \ref{fig:kinannCS}). 
One finds that in the limit $Q_T \to 0$: $\hat{z}_{CS}$ and $\hat{z}_{GJ}$
coincide and also $\hat{x}_{CS}$ and $\hat{x}_{GJ}$. When $Q_T \neq 0$ they
differ only by a rotation: 
\ba 
\hat{\bm{z}}_{GJ} = \cos \beta \; \hat{\bm{z}}_{CS} + \sin \beta \;
\hat{\bm{x}}_{CS},\\
\hat{\bm{x}}_{GJ} = - \sin \beta \; \hat{\bm{z}}_{CS} + \cos \beta \;
\hat{\bm{x}}_{CS}, 
\ea 
where 
\beq 
\cos \beta = \frac{Q}{\tilde{Q}},
\quad \sin \beta = \frac{Q_T}{\tilde{Q}}.  
\eeq
We will also refer to the Collins-Soper frame vectors $t,z,x,y$ 
as a perpendicular basis. 

In the cross sections one will encounter the following functions of $y = P_2
\cdot l/ P_2 \cdot q \approx l^-/q^-$, which in the lepton-pair center of mass
frame equals $y=(1 + \cos \theta^*)/2$, where $\theta^*$ is the angle of 
$\hat{\bm{z}}$ with respect to the momentum of the incoming lepton $\bm{l}$:
\begin{eqnarray}
A(y) &=& \left(\frac{1}{2}-y+y^2\right)\ \stackrel{cm}{=} 
\ \frac{1}{4}\, \left( 1 + \cos^2\theta^* \right),\\
B(y) &=& y\,(1-y)\ \stackrel{cm}{=} \ \frac{1}{4}\, \sin^2 \theta^*,\\
C(y) &=& (1-2y) \ \stackrel{cm}{=} \ - \cos \theta^*,\\
D(y) &=& \sqrt{y\,(1-y)}\ \stackrel{cm}{=} \ \frac{1}{2}\, \sin \theta^*.
\end{eqnarray}
In the Gottfried-Jackson frame (Fig.\ \ref{fig:kinannGJ}) 
$\theta^*$ is called $\theta_2$ and in the Collins-Soper frame $\theta$ (Fig.\
\ref{fig:kinannCS}).

The cross sections are obtained from the hadron tensor after contraction with
the lepton tensor. The lepton tensor for unpolarized leptons expressed in the
lepton center of mass is given by ($\hat l_\perp^\mu
= l_\perp^\mu /(Q \sqrt{y(1-y)})$):
\begin{eqnarray}
L^{\mu \nu} & = &Q^2 \Biggl[
- 2 A(y) g_\perp^{\mu \nu} + 4B(y) \hat z^\mu \hat z^\nu
-4B(y)\left( \hat l_\perp^\mu\hat l_\perp^\nu +\frac{1}{2}\,g_\perp^{\mu \nu}
\right) - 2C(y)D(y)\,\hat z_\ph^{\{ \mu}\hat l_\perp^{\nu \}} \Biggr].
\end{eqnarray}
For later use, the contractions of specific tensor structures in the hadron 
tensor are given in Table~\ref{contractions}.
\begin{table}[htb]
\caption{\label{contractions}
Contractions of the lepton tensor $L_{\mu \nu}$ with tensor
structures appearing in the hadron tensor.}
\begin{center}
\begin{tabular}{ccl}
\\
\hline
\\[-3 mm]
$w^{\mu \nu}$ & {} \hspace{3 cm} {} & $L_{\mu \nu} w^{\mu \nu}/(4 Q^2)$ 
\\[2mm] \hline \\[-4 mm] \hline
\\[-2 mm]
$-g_\perp^{\mu \nu}$ & &
$ \left( \frac{1}{2} - y + y^2 \right)$ \\[2mm] 
$a_\perp^{\,\{ \mu} b_\perp^{\nu\}} -
(a_\perp \cdot b_\perp)\,g_\perp^{\mu\nu}$ & &
$- y \left( 1 - y \right)
\vert \bm{a}_\perp \vert \, \vert \bm{b}_\perp \vert\,\cos(\phi_a+\phi_b)$
\\[2mm] 
$\hat z^{\,\{\mu}a_\perp^{\nu\}}$ & &
$-( 1 - 2y)\sqrt{y(1-y)}\,\vert \bm{a}_\perp \vert\,\cos \phi_a$ \\[2mm]
\hline
\end{tabular}
\end{center}
\end{table}

In the $e^+ e^-$ center of mass frame $d^3P_1 d^3 P_2/P_1^0 P_2^0 = (dz_1/z_1)
(z_2Q^2d z_2/4) d^2 \bm{P}_{1\perp} d \Omega_2$, such that 
\begin{equation}
\frac{d\sigma^{(e^+e^-)}}{dz_1 dz_2 d\Omega d^2 \bm{q}_T}
=\frac{\alpha^2}{16\,Q^4} z_1 z_2 \, L_{\mu\nu}{\cal W}^{\mu\nu},
\label{cross3}
\end{equation}
where $d\Omega$ = $2dy\,d\phi^\ell$, with $\phi^\ell$ giving the 
orientation of $\hat{\bm{l}}_\perp$. 

\subsection{\label{sec:SF}Structure functions}

The hadron tensor ${\cal W}^{\mu\nu}$ can be expanded in terms of independent
Lorentz structures which leads to a parameterization in terms of 
structure functions $W_i$. Ignoring lepton polarization and $\gamma$-$Z$
interference, the most general decomposition consists of four
structure functions. Due to the similarity of the process $e^+e^- \to h_1 \,
h_2\, X$ with the Drell-Yan process, we will employ similar notation
here as used for the latter process, i.e.\ we follow the
notation of Lam \& Tung \cite{Lam-78,Lam-80}, Collins \cite{Collins:1978yt}, 
and Argyres \& Lam \cite{AL-82}. 

The structure functions in 
the leptonic center of mass frame 
(or rather the perpendicular basis) are defined as 
\beq
{\cal W}^{\mu \nu} = - g_\perp^{\mu \nu} W_T + \hat{z}^\mu
\hat{z}^\nu W_L - \hat{z}^{\{\mu} \hat{x}^{\nu\}} W_\Delta
- (\hat{x}^{\{\mu} \hat{x}^{\nu\}}
- \hat{x}^2 g_\perp^{\mu \nu}) W_{\Delta
\Delta},
\eeq
such that ${\cal W}^\mu{}_\mu = -(2W_T + W_L)$. 
The structure functions $W_{T,L,\Delta,\Delta\Delta}$
are associated with specific polarizations of the photon~\cite{Lam-78}:
$W_T=W^{1,1}$, $W_L=W^{0,0}$, $W_\Delta=(W^{0,1}+W^{1,0})/\sqrt{2}$, 
and $W_{\Delta \Delta}=W^{1,-1}$, where the first and second superscripts denote
the photon helicity in the amplitude and its complex conjugate, respectively. 
In terms of these structure functions the cross section becomes
\beq
\frac{d\sigma (e^+e^-\to h_1h_2X)}{dz_1 
dz_2 d\Omega d^2{\bm q_T^{}}}=
\frac{3\alpha^2}{4 Q^2}\;z_1^2z_2^2\;\left\{ 
W_T(1+ \cos^2 \theta^*) + W_L (1- \cos^2\theta^*)
+ W_\Delta \sin 2\theta^* \cos \phi^* + W_{\Delta \Delta} \sin^2 \theta^* 
\cos 2\phi^* \right\},
\eeq
or
\ba
\lefteqn{\frac{dN}{d\Omega} \equiv \left(\frac{d \sigma}{d z_1 d z_2 d^2 \bm{q}_T} 
\right)^{-1} \frac{d \sigma}{dz_1 dz_2 d\Omega^* d^2 \bm{q}_T}  =  
\frac{3}{8\pi} \; \frac{1}{2 W_T + W_L} \; \left[ 
W_T (1+ \cos^2 \theta^*) \right.} \nn\\[2 mm]
&& \qquad \qquad \left.\mbox{} + W_L (1- \cos^2\theta^*) 
+ W_\Delta \sin 2\theta^* \cos \phi^* + W_{\Delta \Delta} \sin^2 \theta^* 
\cos 2\phi^* \right].
\ea
Here $\theta^*, \phi^*$ indicate the polar and azimuthal angle in the $e^+
e^-$ center of mass frame (which for the BELLE experiment is not the lab
frame), such that in the Gottfried-Jackson frame $\theta^* = \theta_2$ and
$\phi^* = \phi_1$, the angles we have used before (see Fig.\
\ref{fig:kinannGJ}). 

Another standard notation for the angular dependences in the Drell-Yan process
can be employed here as well: 
\ba
\frac{dN}{d\Omega} & = & 
\frac{3}{4\pi} \; \frac{1}{\lambda+3} \; \left[ 1+ \lambda \cos^2\theta^*
+ \mu \sin 2\theta^* \cos \phi^* + \frac{\nu}{2} \sin^2 \theta^* 
\cos 2\phi^* \right]. \label{lmnnotation}
\ea
Expressing these parameters in terms of the structure functions $W$ one has:
\beq
\lambda=\frac{W_T-W_L}{W_T+W_L}\; , \;\;\;
\mu=\frac{W_\Delta}{W_T+W_L}\; , \;\;\;
\nu=\frac{2W_{\Delta\Delta}}{W_T+W_L} \; .
\label{lmnWrel}
\eeq

One can transform from the CS frame to the GJ frame by using the following
transformation matrix:
\beq
\left( \begin{array}{c} \lambda \\ \mu \\ \nu \end{array} \right)_{GJ}
= \frac{1}{\Delta_{CS}} \left( \begin{array}{ccc}
1-\frac{1}{2}\rho^2 & - 3 \rho & \frac{3}{4}\rho^2 \\
\rho & 1- \rho^2 & -\frac{1}{2} \rho \\
\rho^2 & 2 \rho & 1+\frac{1}{2} \rho^2 
\end{array} \right) \; 
\left( \begin{array}{c} \lambda \\ \mu \\ \nu \end{array} \right)_{CS},
\label{CStoGJ}
\eeq
where $\rho = Q_T/Q$ and  
\beq
\Delta = 1+ \rho^2 + \frac{1}{2} \rho^2 \lambda 
+ \rho \mu - \frac{1}{4} \rho^2 \nu.
\eeq
The transformation from the GJ frame to the CS frame is the same, but with the
replacement $\rho \to - \rho$. 

In the discussion of the contributions to the various structure functions
given below, the structure function $W_{\Delta \Delta}$ and the 
analyzing power $\nu$ of the $\cos 2\phi$ asymmetry will
receive particular attention due to its Collins effect contributions.   

\section{\label{sec:LOxs}Leading order cross section}

In this section we investigate the $\phi$ dependence that arises in leading
order in $\alpha_s$ and $1/Q$ in the cross section of the process $e^+ e^- \to
h_1 \; h_2 \; X$ differential in the transverse momentum
$\bm{P}_{1\perp}=-z_{1} \bm{q}_T^{}$. The cross section involves products of
fragmentation functions, which unlike the ordinary collinear functions include
transverse momentum dependence \cite{Coll-S-82}. The idea that such
``intrinsic'' transverse momentum will give rise to power suppression turns
out not to be true, even though this was true in the pioneering studies
\cite{Cahn-78,Berger-80} on azimuthal dependences due to intrinsic transverse
momentum. Nontrivial quark spin effects,
which require nonzero partonic transverse momenta, can arise at leading power.
One such effect is the Collins effect, which gives rise to a $\cos 2 \phi $
asymmetry. This was first pointed out in Refs.\ \cite{BJM97,BJM98} and in this
section we will repeat the essentials.

\subsection{\label{sec:Integrated}Integration over transverse photon momentum}
 
\begin{figure}[htb]
\begin{center}
\leavevmode \epsfxsize=6cm \epsfbox{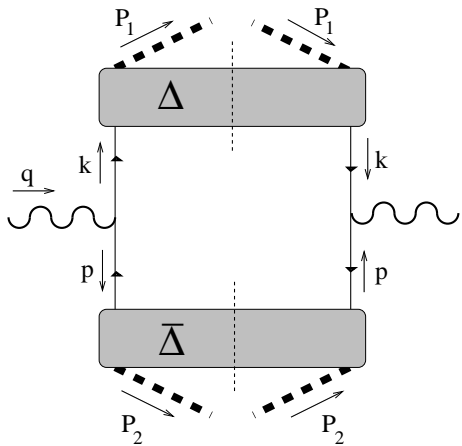}
\caption{\label{Leading} Factorized diagram contributing to $e^+ e^-$ 
annihilation 
in leading order. There is a similar diagram with reversed fermion flow.}
\vspace{-5 mm}
\end{center}
\end{figure}

Although we are interested in the two-hadron inclusive cross section
differential in $\bm{q}_T$, we will first consider the case of integration
over the transverse momentum of the photon. At tree level one needs to
calculate the diagram shown in Fig.\ \ref{Leading}.
It depicts the squared amplitude of the process in which the photon
produces a quark and an antiquark, which subsequently fragment independently 
into the hadrons $h_1$ and $h_2$, respectively.  
The quark fragmentation correlation function $\Delta (P_1;k)$ 
is defined as \cite{Coll-S-82}:
\beq
\Delta_{ij}(P_1;k) = \sum_X \int \frac{d^4x}{(2\pi)^4}\ e^{ik\cdot x} \,
\langle 0 \vert \psi_i(x) \vert P_1; X \rangle
\langle P_1; X \vert \overline \psi_j(0) \vert 0 \rangle, \label{qFcor}
\eeq
where $k$ is the quark momentum and an averaging over color indices is 
left implicit. It is also understood that appropriate path-ordered exponentials
should be included in order to obtain a color gauge invariant quantity, 
cf.\ e.g.\ \cite{Boer:2003cm}. 
In Sec.\ \ref{sec:Univ} we will briefly address the 
universality issue that arises from the proper gauge invariant definition. 

The above matrix element as function of invariants is assumed to vanish
sufficiently fast above a characteristic hadronic scale (${\cal O}(M)$)
which is much smaller than $Q$. This means that in the above matrix
elements $k^2, k \cdot P_1 \ll Q^2$. Hence, we make the following Sudakov
decomposition for the quark momentum $k$:
\begin{equation}
k \equiv \frac{z_{1} Q}{z\sqrt{2}}\,n_-
+ \frac{z (k^2 + \bm{k}_T^2)}{z_{1} Q \sqrt{2}}\,n_+ + k_T^{}
 \approx  \frac{1}{z} P_1 + k_T^{}.
\end{equation}

The Dirac structure of the quark correlation function can be expanded in a
number of amplitudes, i.e., functions of invariants built up from the quark
and hadron momenta, constrained by hermiticity and parity. In the calculation
of the cross section integrated over the transverse momentum of the photon, at
leading order we only encounter the integrated correlation function $\int
dk^+\, d^2 \bm{k}_T \Delta (P_1;k)$, which is a function of $k^-$ only. At
leading twist this leaves only one possible Dirac structure: 
\beq
\Delta(z) \equiv \left. \frac{z}{4} \int dk^+\, d^2 \bm{k}_T\ \Delta(P_1;k)
\right|_{k^- = P_1^-/z} =
\frac{1}{4P_1^-} D_1(z) \, \slsh{\!P_1}.
\label{Deltaexp}
\eeq
The function $D_1(z)$ is the ordinary unpolarized fragmentation function. 

For the fragmentation of an antiquark most things are analogous to the quark
fragmentation. The main difference in the present case is that the role of the
$+$ and the $-$ direction is reversed. We will denote the antiquark correlation
function by $\overline \Delta (P_2;p)$:
\begin{equation}
\overline \Delta_{ij}(P_2;p) = \sum_X 
\int \frac{d^4x}{(2\pi)^4}\ e^{-ip\cdot x}\,
\langle 0 \vert \overline \psi_j(0) \vert P_2; X \rangle
\langle P_2; X \vert \psi_i(x) \vert 0 \rangle.
\end{equation}
Similarly,
\beq
\overline \Delta(\bar z) \equiv \left. \frac{\bar z}{4} \int dp^-\, d^2 
\bm{p}_T\ \Delta(P_2;p)
\right|_{p^+ = P_2^+/{\bar z}} =
\frac{1}{4P_2^+} \overline D{}_1(\bar z) \, \slsh{\!P_2}.
\label{Deltabarexp}
\eeq

The four-momentum conservation delta-function at the photon vertex is written 
as (neglecting $1/Q^2$ contributions)
\begin{equation} 
\delta^4(q-k-p)=\delta(q^+-p^+)\, \delta(q^--k^-)\, \delta^2(
\bm{p}_T^{}+
\bm{k}_T^{}-\bm{q}_T^{}),
\label{deltafn}
\end{equation}
fixing $P_2^+/\bar z=p^+=q^+=P_2^+/z_2$ and $P_1^-/z=k^-=q^-=P_1^-/z_1$. 
Eq.\ (\ref{deltafn}) shows why only the $k^+$ and 
$p^-$-integrated correlation functions are relevant. 
  
The hadron tensor as function of $\bm{q}_T$ is given by 
\beq
{\cal W}^{\mu\nu}=3 \int dp^- dk^+ d^2\bm{p}_T^{} d^2 
\bm{k}_T^{}\, \delta^2(\bm{p}_T^{}+
\bm{k}_T^{}-\bm{q}_T^{})\, 
\text{Tr}\left( \overline \Delta (p) 
\gamma^\mu \Delta 
(k) \gamma^\nu \right). 
\label{WmunuD}
\eeq
The factor 3 originates from the color summation.
We have omitted the flavor indices and summation; furthermore, there is a 
contribution from diagrams with reversed fermion flow, which results from the 
above expression by replacing $\mu \leftrightarrow \nu$ and $q \rightarrow
-q$ (and in the end in a summation over flavors and antiflavors). Note that 
the quark and antiquark transverse
momentum integrations are linked, unless one integrates over $\bm{q}_T$. 

After integration over the transverse momentum of the photon (or equivalently 
over the perpendicular momentum of hadron 1: $\bm{P}_{1\perp}=-z_{1} 
\bm{q}_T^{}$),
the integrations over $\bm{k}_T^{}$ and 
$\bm{p}_T^{}$ in the hadron tensor in Eq.\ 
(\ref{WmunuD}) can be performed resulting in\footnote{Eq.\ (\ref{dqtW}) 
corrects Eq.\ (75) of Ref.\ \cite{BJM97} by an additional factor 
$(z_1 z_2)^{-2}$.}
\beq
\int d^2{\bm q_T^{}}\;{\cal W}^{\mu\nu}
 =  - \frac{12}{z_1 z_2} 
\sum_{a,\bar a}e_a^2 \, g_\perp^{\mu\nu} D_1\overline D_1,
\label{dqtW}
\eeq
We have now included the summation over flavor indices and $e_a$ is the quark
charge in units of $e$. 
The fragmentation functions are flavor dependent and only depend on the
longitudinal momentum fractions, i.e.\ $D_1\overline D_1$ 
= $D_1^a(z_1)\,\overline{D}{}_1^a(z_2)$.

From this hadron tensor one arrives at the following expression for the
cross section at leading order in $\alpha_s$ and $1/Q$
\begin{eqnarray} 
\frac{d\sigma(e^+e^-\to h_1h_2X)}{dz_1 dz_2 d\Omega} & = & 
\frac{3\alpha^2}{Q^2}\; A(y)\; \sum_{a,\bar a}e_a^2\;
                 D_1\overline D_1.
\end{eqnarray} 

\subsection{\label{sec:Unintegrated}Unintegrated cross section}

Now we turn to the cross section differential in the transverse momentum. 
In this case the correlation function $\Delta$ only integrated over $k^+$ 
needs to be considered. It can be 
parameterized in terms of transverse momentum dependent (TMD) 
fragmentation functions \cite{Levelt:1994np,Mulders-Tangerman-96}
\beq
\Delta(z,\bkt) \equiv \left. \frac{1}{4z} \int dk^+\ \Delta(P_1;k)
\right|_{k^- = P_1^-/z,\ \bm{k}_{\scriptscriptstyle T}} =
\frac{M_1}{4P_1^-} \Biggl\{
D_1(z ,\bm{k}_T^2)\,\frac{\slsh{\!P_1}}{M_1}
+ H_{1}^\perp(z ,\bm{k}_T^2)\,\frac{\sigma_{\mu \nu} k_T^\mu P_1^\nu}{M_1^2}
\Biggr\},
\label{DeltaexpB}
\eeq
where we only display the fragmentation functions that are relevant for
unpolarized hadron production. The functions $D_1$ and $H_1^\perp$ yield
contributions to the cross section that are of leading order in $1/Q$.  After
integration over $\bm{k}_T$ the term with $H_1^\perp$ drops out and the first
term reduces\footnote{The relation between TMD and ordinary collinear
  fragmentation functions is not trivial beyond leading order in
  $\alpha_s$. For a discussion on the analogous problem for
  distribution functions we refer to Ref.\
  \cite{Collins:2003fm,Bacchetta:2008xw}.}  
to the expression in Eq.\ (\ref{Deltaexp}). Strictly speaking,
the TMD fragmentation functions depend on $z$ (the lightcone momentum fraction
$z = P_1^-/k^-$ of the produced hadron with respect to the fragmenting quark)
and on $\bm{k}_T^\prime{}^2 = z^2 \bm{k}_T^2$. Here $\bm{k}_T^\prime \equiv
-z\bm{k}_T^{}$ is the transverse momentum of the hadron in a frame where the
quark has no transverse momentum. In order to switch from quark to hadron
transverse momentum a Lorentz transformation leaving $k^-$ and $P_1^-$
unchanged needs to be performed \cite{Coll-S-82,Levelt-Mulders-94}.

The so-called Collins effect function $H_1^\perp$ implies a correlation
between the transverse polarization direction of the quark and the transverse
momentum direction of the unpolarized hadron it fragments into
\cite{Collins-93b}. The Collins effect correlates the azimuthal angle of the
transverse spin of a fragmenting quark with that of the transverse momentum of
the produced hadron (both taken around the quark momentum), via a $\sin
\phi$ distribution of their difference angle $\phi$. Therefore, the
distribution of final state particles contains information about the spin
direction of fragmenting quarks. In this sense it is the strong interaction
analogue of the self-analyzing property of weak decays.

The presence of the Dirac matrix $\sigma_{\mu \nu}$ shows that the Collins
effect is a chiral-odd (a quark-chirality flip) state; an interference term
between opposite chirality states of the fragmenting quark.  The function
$H_1^\perp$ is also often referred to as `time-reversal odd' fragmentation
function, due to its behavior under time reversal. It does not imply a
violation of time reversal symmetry though. For a detailed discussion cf.\
Ref.\ \cite{Boer:2003cm}.

The $\cos 2 \phi$ asymmetry to be discussed below depends on a product of two
Collins effect fragmentation functions $H_1^\perp$. It is an azimuthal {\em
  spin\/} asymmetry in the sense that the asymmetry arises due to the
correlation of the transverse spin states of the quark-antiquark pair. On
average the quark and antiquark will not be transversely polarized, but for
each particular event the spins can have a transverse component and these
components will be correlated via the photon polarization state, which in turn
is determined by the lepton direction. Due to the Collins effect the
directions of the produced hadrons are correlated to the quark and antiquark
spin and hence, to the lepton direction. This correlation does not average out
after summing over all quark polarization states. 

As a transverse spin state is a helicity-flip state, one deduces that the
asymmetry arises from the interference between the photon helicity $\pm 1$
states (along the quark-antiquark axis) and hence contributes to $W_{\Delta
  \Delta}$ and to $\nu$. Such a helicity-flip contribution can also arise from
quark mass terms, but those are power suppressed and do not lead to an
azimuthal dependence (the $\theta$ dependence will be identical though).

The Collins effect was the main reason for studying
azimuthal asymmetries in the BELLE data, since it shows up at
leading order (in both $\alpha_s$ and $1/Q$) in an azimuthal $\cos 2\phi$
asymmetry in the differential cross section for 
unpolarized $e^+ \, e^- \to h_1 \, h_2\, X$ \cite{BJM97}:
\ba
\lefteqn{\frac{d\sigma (e^+e^-\to h_1h_2X)}{dz_1 
dz_2 d\Omega d^2{\bm q_T^{}}}=
\frac{3\alpha^2}{Q^2}\;z_1^2z_2^2\;\Bigg\{ 
          A(y)\;{\cal F}\left[D_1\overline D_1\right]} \nn\\
&&\qquad \qquad \qquad + B(y)\;\cos 2\phi_1 \;
             {\cal F}\left[\left(2\,\hat{\bm{h}}\!\cdot \!
\bm k_T^{}\,\,\hat{\bm{h}}\!\cdot \!
\bm p_T^{}\,
                    -\,\bm k_T^{}\!\cdot \!
\bm p_T^{}\,\right)
                    \frac{H_1^{\perp}\overline H_1^{\perp}}{M_1M_2}\right]
\Bigg\},
\label{LO-OOO}
\ea
where we use the convolution notation 
\begin{equation} 
{\cal F}\left[D\overline D\, \right]\equiv\sum_{a,\bar a}e_a^2\;
\int d^2\bm k_T^{}\; d^2\bm p_T^{}\;
\delta^2 (\bm p_T^{}+\bm k_T^{}-\bm 
q_T^{})  D^a(z_{1},z_{1}^2 \bm{k}_T^2) 
\overline D^a(z_{2},z_{2}^2 \bm{p}_T^2).
\end{equation}
The angle $\phi_1$ is the azimuthal angle 
of $\hat{\bm{h}}=\hat{\bm{x}}$, see Fig.~\ref{fig:kinannGJ}.
So we find that in the lepton-pair center of mass frame:
\begin{eqnarray} 
\lefteqn{
\frac{dN}{d\Omega} \equiv \left(\frac{d \sigma}{d z_1 d z_2 d^2 \bm{q}_T} 
\right)^{-1} \frac{d \sigma}{dz_1 dz_2 d\Omega d^2 \bm{q}_T} =  
\frac{3}{16 \pi} \;\Bigg\{ (1+ \cos^2\theta_2) 
          {\cal F}\left[D_1\overline D_1\right]} \nonumber \\
&&\qquad \mbox{} + \sin^2 \theta_2 \;\cos 2\phi_1 \;
             {\cal F}\left[\left(2\,\hat{\bm{h}}\!\cdot \!
\bm k_T^{}\,\,\hat{\bm{h}}\!\cdot \!
\bm p_T^{}\,
                    -\,\bm k_T^{}\!\cdot \!
\bm p_T^{}\,\right)
                    \frac{H_1^{\perp}\overline H_1^{\perp}}{M_1M_2}
\right] \Bigg\}\bigg/ {\cal F}\left[D_1\overline D_1\right].
\label{reducedOO}
\end{eqnarray}
This shows that
\ba
W_T & = & {\cal F}\left[D_1\overline D_1\right],\\
W_L & = & W_\Delta \ = \ 0,\\
W_{\Delta \Delta} & = & {\cal F}\left[\left(2\,\hat{\bm{h}}\!\cdot \!
\bm k_T^{}\,\,\hat{\bm{h}}\!\cdot \!
\bm p_T^{}\,
                    -\,\bm k_T^{}\!\cdot \!
\bm p_T^{}\,\right)
                    \frac{H_1^{\perp}\overline H_1^{\perp}}{M_1M_2}
\right],
\ea
or equivalently, that at tree level $\lambda=1, \mu=0$ and
\beq
\nu = 2 \frac{ {\cal F}\left[\left(2\,\hat{\bm{h}}\!\cdot \!
\bm k_T^{}\,\,\hat{\bm{h}}\!\cdot \! \bm p_T^{}\, -\,\bm k_T^{}\!\cdot \!
\bm p_T^{}\,\right) H_1^{\perp}\overline H_1^{\perp} 
\right]}{M_1M_2{\cal F}\left[D_1\overline D_1\right]}.
\label{nufromCollins}
\eeq 
We emphasize that measuring $\nu$ does not involve a measurement of
the polarization of the produced hadrons nor of the incoming leptons.  
Also, the result is not suppressed by factors of $1/Q$, in contrast to
the $\cos 2\phi $
asymmetry discussed by Berger \cite{Berger-80}, which is $1/Q^2$-suppressed.
Eq.\ (\ref{nufromCollins}) applies to the GJ frame, but upon neglecting
$Q_T^2/Q^2$ power suppressed terms it is the same expression in the CS frame, 
as can be seen using Eq.\ (\ref{CStoGJ}). 

We will often assume Gaussian 
$\bm{k}_T^{}$-dependence of the various functions, since in that case 
the convolutions can be explicitly 
evaluated. Eq.\ (\ref{LO-OOO}) then becomes
\ba
\lefteqn{\frac{d\sigma (e^+e^-\to h_1h_2X)}{dz_1 
dz_2 d\Omega d^2{\bm q_T^{}}}=
\frac{3\alpha^2}{Q^2}\;{\cal G}(Q_T^{}; R) \sum_{a,\bar a}e_a^2\;\Bigg\{ 
          A(y)\;D_1^a(z_1) 
\overline{D}{}_1^a(z_2)} \nn\\
&&\qquad \qquad \qquad + B(y)\;\cos 2\phi_1 \;
   \frac{Q_T^{2} R^4}{M_1M_2 R_1^2 R_2^2} H_1^{\perp a}(z_1) 
\overline{H}{}_1^{\perp a}(z_2) \Bigg\},
\label{LO-OOOB}
\ea
where $R^2=R_1^2 R_2^2 /(R_1^2 + R_2^2)$ and 
\beq
D_1(z_1,
\bm{k}_T^\prime{}^2) = D_1(z_1) R_1^2 
\exp(-R_1^2 \bm{k}_T^2) / \pi z_1^2 \equiv 
D_1(z_1) {\cal G}(|\bm{k}_T^{} |; R_1)/ z_1^2,
\eeq
and similarly for $\overline D_1, H_1^\perp, \overline H{}_1^\perp$ with
obvious replacements (for details cf.\ Ref.\ 
\cite{Mulders-Tangerman-96}). For simplicity we have assumed the same Gaussian
width for $D_1$ and $H_1^{\perp}$, which is not expected to be realistic. 
Later on we
will drop this assumption. Rather, we will often assume $R_1=R_2=R$
which means equal widths for $H_1^\perp$ and $\overline{H}{}_1^\perp$, 
$R_{1u}=R_{2u}=R_u$ which means equal widths for $D_1$ and
$\overline{D}{}_1^{}$, and moreover, $M_1=M_2=M$, which should be a
reasonable assumption when the two produced hadrons are charged
pions. This leads to 
\ba
\lefteqn{\frac{d\sigma (e^+e^-\to h_1h_2X)}{dz_1 
dz_2 d\Omega d^2{\bm q_T^{}}}=
\frac{3\alpha^2}{Q^2}\; \sum_{a,\bar a}e_a^2\;\Bigg\{ 
          A(y)\; {\cal G}(Q_T^{}; R_u/2) \;D_1^a(z_1) 
\overline{D}{}_1^a(z_2)}  \nn\\
&&\qquad \qquad \qquad + B(y)\;\cos 2\phi_1 \;
   \frac{Q_T^{2}}{4M^2} \; {\cal G}(Q_T^{}; R/2) \; H_1^{\perp a}(z_1) 
\overline{H}{}_1^{\perp a}(z_2) \Bigg\}.
\label{LO-OOOC}
\ea

\subsection{\label{sec:Weigh}Weighted cross sections}

The expressions in the previous subsection contain convolutions,
which are not the objects of interest, rather one wants to learn about 
the fragmentation functions depending on $z$ and $\bm{k}_T^2$. This may not be
possible without further assumptions about the type of $\bm{k}_T^2$
dependence, such as assuming Gaussian transverse momentum dependence. 
As a way out, it has been suggested
\cite{Kotzinian:1995cz,BJM97} to consider 
specific integrated, weighted asymmetries that probe instead of the 
full transverse momentum dependence, the 
$\bm{k}_T^2$-moments of the functions. These so-called transverse moments are 
defined as:
\begin{equation}
F^{(n)}(z_{1})= \int d^2 \bm{k}_T^\prime \, \left(
\frac{\bm{k}_T^2}{2 M_1^2}\right)^n 
F(z_{1},\bm{k}_T^\prime{}^2),
\end{equation}
for a generic fragmentation function $F$. 
In particular, the first transverse moment of the Collins fragmentation
function  
\beq 
H_1^{\perp (1)}(z) = z^2 \int d^2 \bkt \, \frac{\bm{k}_T^2}{2M^2}\, 
H_1^\perp (z,z^2\, \bm{k}_T^2)
\eeq
has been considered frequently in the literature. 

In Sect.\ \ref{sec:Integrated} we have presented the hadron tensor and
cross section integrated over transverse momentum of the photon.  A
number of structures averaged out to zero, which are retained when the
integration is weighted with an appropriate number of factors of
$\bm{q}_T^{}$. By constructing such weighted cross sections {\em at
  tree level\/} the convolutions become simply products of such 
$\bm{k}_T^2$-moments. The $\bm{k}_T^2$-moments can be used in other processes 
where they also occur. 

To shorten the notation we define weighted cross sections as follows
\begin{equation}
\left< W \right>=
\int d^2\bm q_T^{}\, W\,\frac{d\sigma (e^+e^-\to h_1h_2X)}{dz_1 dz_2 d\Omega 
d^2{\bm q_T^{}}},
\end{equation}
where $W$ = $W(Q_T,\phi_1)$. 
In this way one finds:
\ba
\left< 1 \right> 
\ & = &  \ 
\frac{3\,\alpha^2}{Q^2}\; A(y)\;  
\sum_{a,\bar a} e_a^2\;D_1^a(z_1)\,\overline D{}_1^a(z_2),
\label{wXsection-1_O}\\[2 mm]
\left< Q_T^2 \right> 
\ & = & \ 
\frac{3\,\alpha^2}{Q^2}\; 2 A(y)\;
\sum_{a,\bar a}e_a^2\; \left(
    {M_1}^2D_1^{(1)a}(z_1)\overline D{}_1^a(z_2) +
    {M_2}^2D_1^a(z_1)\overline D{}_1^{(1)a}(z_2) \right), 
\label{wXsection-QT_O}\\[2 mm]
\left< \frac{Q_T^2}{4 M_1M_2}\,\cos 2\phi_1  \right> 
\ & = & \ 
\frac{3\,\alpha^2}{Q^2}\; B(y)\;
\sum_{a,\bar a}e_a^2\; 
\;H_1^{\perp(1)a}(z_1)
\,\overline H_1^{\perp (1)a}(z_2).
\label{wXsection-cos2phi_O}
\ea
The $\bm{k}_T^2$-moment $H_1^{\perp
  (1)}$ that arises in the above $e^+e^-$-annihilation expression also appears 
in the $Q_T$-weighted $\sin(\phi_h+\phi_S)$ asymmetry in 
semi-inclusive lepton-hadron scattering, in that case
multiplied by the transversity distribution function \cite{BM98}.
This illustrates the purpose of considering such weighted cross sections. 

Next we will discuss an estimate of the tree level 
weighted expressions of interest for the Collins effect. 
In Sec.\ \ref{sec:Weigh2} 
we will address the effect of radiative corrections on
such weighted asymmetries. 

\subsection{\label{sec:Estimate}Estimate of the $Q_T^2$-weighted Collins effect asymmetry}

A natural question to ask is what one expects for the magnitude of the
Collins effect asymmetry. This runs immediately into the problem that the
Collins fragmentation function is a nonperturbative quantity that is at least
as hard to calculate from first principles as the ordinary unpolarized
fragmentation function $D_1$. In this subsection we
will discuss a rough estimate of the weighted cross section
defined in Eq.~(\ref{wXsection-cos2phi_O}), appropriately normalized. 

For an order of magnitude estimate of the weighted asymmetry, we
consider the situation of the produced hadrons being a $\pi^+$ and a
$\pi^-$ and only consider up and down quarks. 
Furthermore, we assume $D_1^{u\to\pi^+}(z)=D_1^{\bar d\to\pi^+}(z)$,
$D_1^{d\to\pi^-}(z)=D_1^{\bar u\to\pi^-}(z)$ and 
neglect unfavored
fragmentation functions like $D^{d\to\pi^+}(z)$, etc; and similarly
for the time-reversal odd functions $H_1^\perp, \overline{H}{}_1^\perp$.
These equalities 
seem quite safe for the $D_1$ functions on grounds of isospin and charge 
conjugation. The same
assumptions might be non-trivial for the $H_1^\perp$ functions and remain to
be tested. Studies of the HERMES data seem to indicate that the unfavored
Collins functions can be of the same magnitude as the favored ones
\cite{Airapetian:2004tw}.  It remains to be seen whether this also holds true
at higher energy scales, when the multiplicity of hadrons in the final state
is significantly higher and momentum conservation would not correlate
the hadron momenta as much as at low multiplicities. 

With these assumptions we obtain
\begin{equation}
\left< \frac{Q_T^2}{4M_{\pi}^2}\,\cos 2\phi_1  \right>
= F(y)
\,\frac{H_1^{\perp(1)}(z_1)}{D_1(z_1)}
\,\frac{H_1^{\perp(1)}(z_2)}{D_1(z_2)}
\,\left< 1 \right>,
\label{ratio}
\end{equation}
where
\begin{equation}
F(y)= \frac{B(y)}{A(y)}\stackrel{cm}{=} \frac{\sin^2 \theta_2}{1 
+ \cos^2\theta_2}. 
\label{factor}
\end{equation}
An upper bound on the ratio $H_1^{\perp(1)}(z_1)/D_1(z_1)$ could be
obtained using the Soffer type of bound 
\cite{Bacchetta:1999kz,Bacchetta:2002tk} 
\beq
|\bm{k}_T^{}| \; \left| H_1^{\perp}(z,|\bm{k}_T^{}|) \right| \leq M_h \; 
D_1(z,|\bm{k}_T^{}|),
\label{bound}
\eeq
which should hold for all $|\bm{k}_T^{}|$. It implies: 
\beq
\frac{|H_1^{\perp (1)}(z)|}{D_1(z)} \leq \amp{|\bm{k}_T^{}|}(z)/(2 M_h).
\label{bound2}
\eeq 
Alternatively, several
model calculations of the Collins function have been performed and could 
be used to obtain an estimate of the weighted asymmetry, rather than of an
upper bound on it. A review of models has been
presented in Ref.\ \cite{Amrath:2005gv}. The first model was given by Collins 
\cite{Collins-93b} and employed in Ref.\ \cite{Kotzinian-Mulders-97} to 
estimate the Collins effect asymmetry in SIDIS, which led to the conclusion 
that $H_1^{\perp(1)}(z_1)/D_1(z_1) ={\cal O} (1)$. Here we will use
this result and not worry about the sign of the Collins function. 
Also, we will use that the average value of $F(y)$ is approximately
0.5. 

In order to get an estimate of the true asymmetry without artificial
enhancement of the weight, one should compare Eq.~(\ref{wXsection-cos2phi_O}) 
with the weighted cross section $\left< Q_T^2/4M_{\pi}^2 \right>$, rather than
with $\left< 1 \right>$. From Eqs.\ (\ref{wXsection-1_O}) and 
(\ref{wXsection-QT_O}) one obtains:
\beq
 \left< \frac{Q_T^2}{4M_\pi^2} \right>
= \frac{1}{2} \left(\frac{D_1^{\perp(1)}(z_1)}{D_1(z_1)}
+ \frac{D_1^{\perp(1)}(z_2)}{D_1(z_2)} \right)
\,\left< 1 \right>.
\eeq
This we estimate by using that $D_1^{(1)}/D_1 = 
\amp{\bm{k}_T^{\prime}{}^2}(z)/(2z^2M^2)$. Ref.\
\cite{Anselmino:1999pw} presented a fit to LEP data to find
that the average transverse momentum squared of pions inside a jet can
be parameterized well as 
\beq
\amp{\bm{k}_T^{\prime}{}^2}^{\half}(z) = 0.61 \, z^{0.27} (1-z)^{0.20} \ 
\text{GeV}/c.
\label{avgkt}
\eeq
This leads at $z_1=z_2=1/2$ (where the average transverse momentum
squared is maximal approximately) to 
$\left< Q_T^2/4M_\pi^2 \right> \approx 20 \,\left< 1 \right>$ 
and hence to an estimate at the few percent level for the ratio 
$\left<\left(Q_T^2/4M_\pi^2\right) \,\cos 2\phi_1  \right>
/\left<Q_T^2/4M_\pi^2 \right>$. Of course, this should not be viewed
as more than a crude estimate. 

Note that $H_1^{\perp(1)}(z_1)/D_1(z_1) ={\cal O} (1)$ is consistent with the
bound in Eq.\ (\ref{bound2}), if one uses the maximum of 
$\amp{\bm{k}_T^{\prime}{}^2}^{\half}(z) \approx 0.44$ (around $z=1/2$) for 
$z \amp{|\bm{k}_T^{}|}(z)$, which leads to $H_1^{\perp(1)}(z_1)/D_1(z_1) 
\simorderr 3$. 

Another model study 
\cite{Bacchetta:2002tk} has also 
obtained a prediction for the twice-weighted asymmetry (cf.\ Eqs.\
(\ref{wXsection-QT_O}) and (\ref{wXsection-cos2phi_O}) with $M_1=M_2$) 
\ba
\amp{P_{h\perp}^2 \cos 2\phi_1}(\theta_2,z_1,z_2) & = & \frac{\int d^2
  \bm{P}_{h\perp} P_{h\perp}^2 \cos 2\phi_1 \,d^5 \sigma}{\int d^2
  \bm{P}_{h\perp} P_{h\perp}^2 d^5 \sigma } \nn\\
& = & \frac{2B(y)\;
         H_1^{\perp(1)}\overline H_1^{\perp(1)}}{A(y)\;\bigg(
            D_1\overline D_1^{(1)}
          + D_1^{(1)}\overline D_1
         \bigg)},
\ea
in $e^+e^- \to h_1 \, h_2\, X$. 
The calculation is based on the Manohar-Georgi model \cite{Manohar:1983md}
and reproduces the unpolarized fragmentation function reasonably well. 
It leads for fixed $z_2$ bins to an asymmetry that is almost linearly rising
as function of $z_1$. 
For $0.5 \leq z_2 \leq 0.8$ and $z_1 \sim 0.5$ the weighted asymmetry is
3-4\%. This is similar in size to the estimate obtained above.
This indicates that measuring the weighted asymmetry  
should be feasible for the present-day high luminosity 
electron-positron scattering experiments.

\section{\label{sec:Univ}Universality of the Collins effect}

Applying the results for the Collins function from BELLE in the Collins
asymmetry in semi-inclusive DIS, assumes the Collins function is universal, 
i.e.\ that it is the same for all processes in which it occurs. 
This assumption is not as obvious as it may
seem. For T-odd {\em distribution\/} functions it has been shown 
\cite{Collins:2002kn,Belitsky:2002sm} that they are
process dependent. This follows from their gauge invariant definition as
matrix elements of operators involving path-ordered exponentials that are
nonlocal off the lightcone. 
For T-odd fragmentation functions, such as the Collins
fragmentation function, a similar conclusion was drawn 
\cite{Boer:2003cm,Bomhof:2006ra}. However, for
the particular case of 
$e^+ e^- \to h_1 h_2 X$ and SIDIS, it has been argued
using a model, but later also with more general arguments, that the Collins 
fragmentation functions are identical \cite{Metz:2002iz,Collins:2004nx}.
Recently this conclusion was extended to the process 
$p p \to h\ \text{jet} \ X$ \cite{Yuan:2007nd} and argued to be a generic and
model-independent result. For the
use of the Collins fragmentation functions in hadronic collisions it is
essential though that factorization holds, which is not established however 
\cite{Collins:2007nk,Vogelsang:2007jk,Collins:2007jp,Bomhof:2007xt}. 
A further indication for the universality of the Collins function comes from 
a spectral analysis of the fragmentation
correlation function within a spectator model calculation 
\cite{Gamberg:2008yt}. 

Assuming the universality of the Collins fragmentation function for the
processes $e^+ e^- \to h_1 h_2 X$ and SIDIS, a simultaneous fit to the Collins
effect asymmetry data has been performed and a first
extraction of transversity was obtained \cite{Anselmino:2007fs}. This
demonstrates the feasibility of using the Collins effect to access
transversity and why it is worth doing the Collins effect asymmetry 
measurement at BELLE. For this reason it becomes also important to study
other potential contributions to the asymmetry that arises from the Collins
effect. This we will do in the following sections. 

\section{\label{sec:HT}Higher twist}

\begin{figure}[htb]
\begin{center}
\leavevmode \epsfxsize=12cm \epsfbox{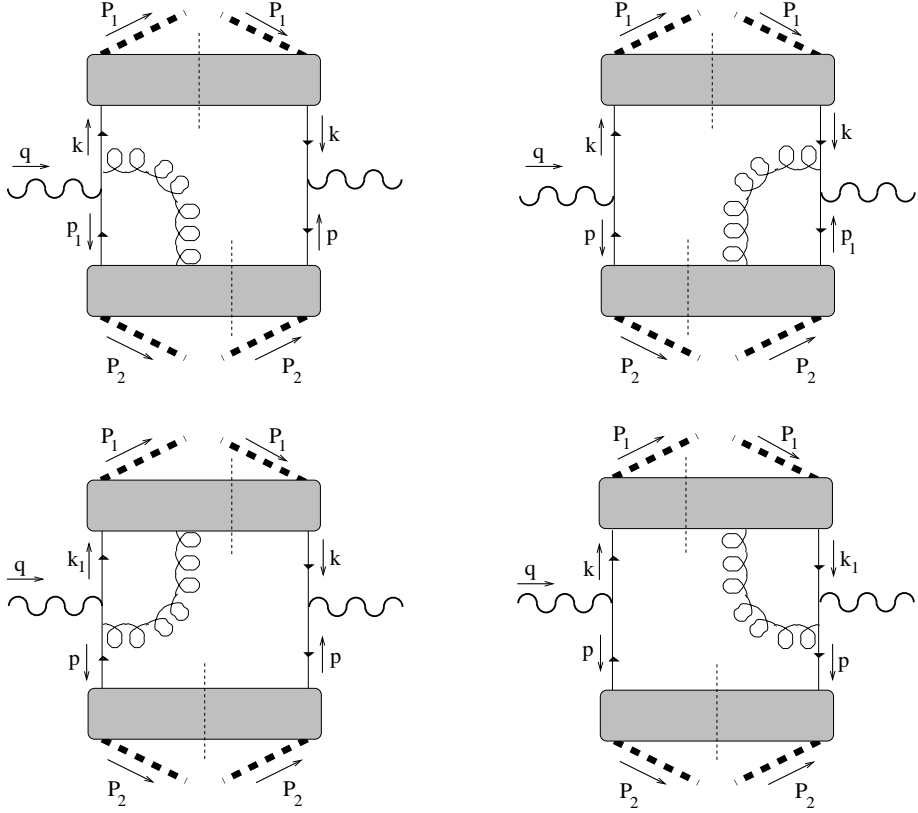}
\caption{\label{Subleading} Diagrams contributing to $e^+ e^-$ 
annihilation at order $1/Q$.}
\vspace{-5 mm}
\end{center}
\end{figure}

In this section we will study the terms that arise when going beyond
leading order in $1/Q$. 
Insertion of the leading order parameterization Eq.\ (\ref{DeltaexpB}) of 
$\Delta$ in
the calculation of the diagram shown in Fig.\ \ref{Leading} also produces 
$1/Q$ contributions. Such 1/Q contribution can already be generated by
simply transforming to a different frame. 
This contribution is not electromagnetically gauge
invariant and the full calculation at order $1/Q$ \cite{BJM97} requires first
of all, that the 
correlation function $\Delta$ is parameterized further to include higher
twist fragmentation functions \cite{Mulders-Tangerman-96} 
\beq
\Delta(z,\bkt) =
\frac{M_1}{4P_1^-} \Biggl\{
D_1\,\frac{\slsh{\!P_1}}{M_1}
+ H_{1}^\perp\,\frac{\sigma_{\mu \nu} k_T^\mu P_1^\nu}{M_1^2}
+ E\, {\bf 1}
+ D^\perp\,\frac{\slsh{k_T^{}}}{M_1}
+ H\,\sigma_{\mu \nu} n_-^\mu n_+^\nu
\Biggr\},
\label{DeltaexpC}
\eeq
and second, inclusion of the diagrams in Fig.\ \ref{Subleading} \cite{Levelt-Mulders-94}.
These four diagrams involve one gluon which connects to one
of the two soft hadronic matrix elements. 

Hence, up to and including 
order $1/Q$ the quark fragmentation is described with help of two types of 
correlation
functions: the quark correlation function $\Delta (P_1;k)$ discussed before
and the
quark-gluon correlation function $\Delta_A^\alpha (P_1;k,k_1)$ 
\cite{Levelt-Mulders-94}:
\beq
\Delta^\alpha_{A\, ij}(P_1; k,k_1) =\sum_X 
 \int \frac{d^4x}{(2\pi)^4}\frac{d^4y}{(2\pi)^4}\ 
e^{ i\,k\cdot y + i\,k_1\cdot(x-y) } \langle 0 \vert
\psi_i(x)\, g A_T^\alpha (y)\, \vert P_1; X \rangle
\langle P_1; X \vert \,\overline \psi_j(0)\vert 0\rangle,
\eeq
where $k, k_1$ are the quark momenta and again inclusion of path-ordered
exponentials and an averaging over color indices are understood. 
Note that the definition of $\Delta_A^\alpha$ includes one power of the
strong coupling constant $g$ and $A_T^\alpha \equiv g_T^{\alpha \beta}
A_{\beta}$. 

In a calculation up to subleading
order, we only encounter the partly integrated correlation functions
$\int dk^+\, \Delta (P_1;k)$ and $\int dk^+  d^4k_1 \,
\Delta_A^\alpha (P_1;k,k_1)$. 
This allows to express the quark-gluon correlation functions in terms of the
quark correlation functions with help of the classical equations of
motion (e.o.m.) \cite{Politzer:1980me}. 
In the subleading terms of the cross section one encounters 
functions indicated with a tilde ($\tilde E$, $\tilde H$, \ldots), which 
differ from the corresponding twist-3 functions ($H$, $E$, \ldots) by a
twist-2 part, namely
\begin{eqnarray}
& & E=\frac{m}{M_1}z D_1 + \tilde{E},\\
& & D^\perp=z D_1 + \tilde{D}^\perp,\\[2 mm]
& & H=-\frac{\bm{k}_T^2}{M_1^2}z H_1^\perp + \tilde{H}.
\end{eqnarray}
For more details we refer to \cite{BJM97}.

The five diagrams lead to the following expression for the hadron tensor 
up to and including order $1/Q$:
\begin{eqnarray}
\lefteqn{{\cal W}^{\mu\nu}=3 \int dp^- dk^+ d^2\bm{p}_T^{} d^2 
\bm{k}_T^{}\, \delta^2(\bm{p}_T^{}+
\bm{k}_T^{}-\bm{q}_T^{})\, \Biggl\{ 
\text{Tr}\left( \overline \Delta (p) 
\gamma^\mu \Delta 
(k) \gamma^\nu \right)} \nonumber \\[3mm]
&&- \text{Tr}\left( \overline \Delta_A^\alpha (p) \gamma^\mu \Delta 
(k) \gamma_\alpha \frac{\slsh{n_+}}{Q\sqrt{2}} \gamma^\nu \right) 
- \text{Tr}\left( \left(\gamma_0 \overline \Delta_A^{\alpha\dagger} (p) 
\gamma_0
\right) \gamma^\mu
 \frac{\slsh{n_+}}{Q\sqrt{2}} \gamma_\alpha \Delta (k) \gamma^\nu \right)
\nonumber \\[3mm]
&& + \left. \text{Tr} \left( \overline \Delta (p) \gamma^\mu \left(\gamma_0
 \Delta_A^{\alpha\dagger} (k) \gamma_0\right) \gamma^\nu 
\frac{\slsh{n_-}}{Q\sqrt{2}} \gamma_\alpha 
\right) 
+ \text{Tr}\left( \overline \Delta(p) \gamma_\alpha
 \frac{\slsh{n_-}}{Q\sqrt{2}} \gamma^\mu \Delta_A^\alpha (k) \gamma^\nu 
\right) \Biggr\} \right|_{p^+,k^-}. \label{Wmunustart}
\end{eqnarray}
The terms with $\slsh{n_\pm}$ arise from the fermion 
propagators in the hard part neglecting contributions that will appear 
suppressed by powers of $Q^2$, 
\begin{eqnarray}
\frac{\slsh{q}-\slsh{p_1}+m}{(q-p_1)^2-m^2} &\approx& 
\frac{(q^+-p_1{}^+)\gamma^-}{2(q^+-p_1{}^+)q^-}=\frac{\gamma^-}{2q^-}
=\frac{\slsh{n_+}}{Q\sqrt{2}}, \label{eeprop1}
\\[3mm]
\frac{\slsh{k_1}-\slsh{q}+m}{(k_1-q)^2-m^2} &\approx& 
\frac{(k_1{}^--q^-)\gamma^+}{-2(k_1{}^--q^-)q^+}=\frac{\gamma^+}{-2q^+}
=-\frac{\slsh{n_-}}{Q\sqrt{2}}.\label{eeprop2}
\end{eqnarray}

As mentioned, one can always integrate 
out one of the momenta
of $\Delta_A^\alpha (k,k_1)$ or $\overline \Delta_A^{\alpha} (p,p_1)$
and apply the e.o.m.\ immediately. 
The quantities $\Delta_A^\alpha (k)$ and $\gamma_0 \Delta_A^{\alpha\dagger}
(k) \gamma_0$ arise from integrating out the second and first 
argument of $\Delta_A^\alpha (k,k_1)$, respectively:
\begin{eqnarray}
\int d^4k_1 \Delta_{A\, ij}^\alpha(k,k_1) & = & 
\sum_X \int \frac{d^4x}{(2\pi)^4}\
e^{i\,k\cdot x} \langle 0 \vert \psi_i(x)\, g A_T^\alpha (x)\,\vert P_1; X 
\rangle 
\langle P_1; X \vert 
\overline \psi_j(0) \vert 0 \rangle = \Delta^\alpha_{A\, ij}(k), \\[2mm]
\int d^4k_1\Delta_{A\, ij}^\alpha(k_1,k) & = & 
\sum_X \int \frac{d^4x}{(2\pi)^4}\ e^{i\,k\cdot x} \langle 0 \vert 
\psi_i(x)\,\vert P_1; X \rangle 
\langle P_1; X \vert \,g A_T^\alpha (0)\,
\overline \psi_j(0) \vert 0 \rangle = 
(\gamma_0\Delta^{\alpha \dagger}_{A}\gamma_0)_{ij}(k),
\end{eqnarray} 
and similarly for $\overline \Delta_A^{\alpha} (p)$ and 
$\gamma_0 \overline \Delta_A^{\alpha\dagger} (p) \gamma_0$.

To obtain the expressions for the symmetric and antisymmetric 
parts of the hadron tensor we expand all vectors in $\Delta, \overline \Delta,
\Delta_A^\alpha$ and 
$\overline{\Delta}{}_A^\alpha$ 
in the perpendicular basis ($\hat t$, $\hat z$ and
$\perp$ directions). In particular, we reexpress the momenta 
$k_T^{}$ and $p_T^{}$ in terms of their
perpendicular parts and a part along $\hat t$ and $\hat z$. For this we
need
\begin{equation}
g_T^{\mu\nu} = g_\perp^{\mu \rho}
g_{T\rho}^{\ \ \nu} 
- \frac{Q_T^{}}{Q}\,(\hat t^\mu + \hat z^\mu)\hat h^\nu.
\end{equation}
We will refer to these perpendicular projections as e.g.\ $k_\perp$ rather
than $k_{T\perp}^{}$. Thus
\begin{equation}
k_\perp^\mu \equiv g_\perp^{\mu\nu}k_{T\nu}^{} = k_T^\mu 
+ \frac{\bm{q}_T^{}\cdot 
\bm{k}_T^{}}{Q}\,(\hat t^\mu + \hat z^\mu),
\end{equation}
and similarly for $p_\perp$. We note that 
for these four vectors the two-component perpendicular 
parts are the same as the two-component transverse parts, i.e., 
$\bm k_\perp = \bm k_T^{}$, etc. 
The full expression for the hadron 
tensor is then
\begin{eqnarray} 
\lefteqn{
{\cal W}^{\mu\nu}=12 z_1z_2
\int d^2\bm k_T^{}\; d^2\bm p_T^{}\;
\delta^2 (\bm p_T^{}+\bm k_T^{}
         -\bm q_T^{})\Bigg\{ 
       - g_\perp^{\mu\nu} 
            D_1\overline D_1 -\frac{k_\perp^{\{\mu}p_\perp^{\nu\}}
              +g_\perp^{\mu\nu}\,\bm k_\perp\!\cdot\!\bm p_\perp\,}{M_1M_2} 
         H_1^{\perp}\overline H_1^{\perp}}
\nonumber\\ && \hspace{46mm}
     {}+ 2\frac{\hat z_\ph^{\{\mu}k_\perp^{\nu\}}}{Q} \bigg[
          \frac{\tilde D^{\perp}}{z_1}\overline D_1
          - \frac{{M_2}}{M_1} H_1^{\perp}\frac{\overline H}{z_2}
         \bigg] - 2\frac{\hat z_\ph^{\{\mu}p_\perp^{\nu\}}}{Q} \bigg[
          D_1\frac{\overline D^{\perp}}{z_2}
          - \frac{{M_1}}{M_2}
            \frac{\tilde H}{z_1}\overline H_1^{\perp}
         \bigg]
\Bigg\}. 
\label{WmunuS}
\end{eqnarray} 

The cross section at leading order in $\alpha_s$ but including twist-3
contributions becomes 
\ba
\lefteqn{\frac{d\sigma (e^+e^-\to h_1h_2X)}{dz_1 
dz_2 d\Omega d^2{\bm q_T^{}}}=
\frac{3\alpha^2}{Q^2}\;z_1^2z_2^2\;\Bigg\{ 
          A(y)\;{\cal F}\left[D_1\overline D_1\right]}
\nn\\
&&\qquad \qquad \qquad + B(y)\;\cos 2\phi_1 \;
             {\cal F}\left[\left(2\,\hat{\bm{h}}\!\cdot \!
\bm k_T^{}\,\,\hat{\bm{h}}\!\cdot \!
\bm p_T^{}\,
                    -\,\bm k_T^{}\!\cdot \!
\bm p_T^{}\,\right)
                    \frac{H_1^{\perp}\overline H_1^{\perp}}{M_1M_2}\right]
\nn\\
&&\qquad \qquad \qquad - C(y)D(y)\; \cos \phi_1 \; \left( 
\frac{M_1}{Q} {\cal F}\left[\frac{\hat{\bm{h}}\!\cdot \! \bm k_T^{}}{M_1} 
\frac{\tilde D^{\perp}}{z_1} \overline D_1 \right]
- \frac{M_2}{Q} {\cal F}\left[\frac{\hat{\bm{h}}\!\cdot \! \bm k_T^{}}{M_1}
H_1^{\perp}\frac{\overline H}{z_2} \right] \right.\nn\\
&&\hspace{5.2 cm} \left.
-\frac{M_2}{Q} {\cal F}\left[\frac{\hat{\bm{h}}\!\cdot \! \bm p_T^{}}{M_2} 
D_1\frac{\overline D^{\perp}}{z_2}\right]
+ \frac{M_1}{Q} {\cal F}\left[\frac{\hat{\bm{h}}\!\cdot \! \bm p_T^{}}{M_2}
\frac{\tilde H}{z_1}\overline H_1^{\perp} \right]
\right) \Bigg\}.
\ea
This is the result in Eq.\ (\ref{LO-OOO}) plus an additional $\cos \phi$ 
asymmetry of order $M/Q$. 
The above expression is given in the GJ frame. When transforming to
the CS frame the $\cos 2 \phi$ asymmetry remains unchanged up to terms of
order $M Q_T/Q^2$, as can be seen from Eq.\ (\ref{CStoGJ}).  

The function $E$ only contributes in the case of polarized electrons at $1/Q$
and for unpolarized electrons at the $1/Q^2$ level. The latter would be the
analogue of the contribution considered by Jaffe \& Ji \cite{Jaffe:1991ra} for
the Drell-Yan process after the replacement $\delta^2 (\bm p_T^{}+\bm k_T^{}
-\bm q_T^{}) \to \delta^2 (\bm q_T^{})$.

We have neglected dynamic twist-4 effects, which are order $M^2/Q^2$
corrections. At BELLE these effects are expected to be at most at the percent
level, but nevertheless could be relevant to include. To investigate this
further theoretically requires an extensive study due to the many possible
sources of $1/Q^2$ effects \cite{EFP,Cahn-78,Berger-80,Jaffe:1991ra}, which
thus far has not been undertaken. One reason for this is that factorization
has not been considered yet in this case (actually not even for the twist-3
case thus far). One particular type of higher twist effect was studied by
Berger \cite{Berger-80} for $e^+ e^- \to \pi X$, which is mostly relevant at
large values of $z_h$, towards the exclusive limit. This is like the higher
twist contributions to the Drell-Yan azimuthal asymmetries studied in Refs.\
\cite{Brandenburg-94,Eskola-94}, which contribute mainly at large $x$.
However, it remains to be seen whether this specific $z_h$ dependence holds
true for all types of higher twist effects. They {\em are\/} 
generally expected to
lead to $\mu > \nu$. Therefore, if $\mu$ is found to be much smaller than
$\nu$, it would be a strong indication that one is not dealing with higher
twist effects. Experimentally one could also test whether such effects are
relevant by allowing in the fits to asymmetries for additional $1/Q^2$
dependence, as has been done in studies of DIS data
\cite{Virchaux:1991jc,Alekhin:1999iq,Leader:2002ni}.

\section{\label{sec:EW}Electroweak interference effects}

In the previous sections we have presented the results of the tree-level
calculation of inclusive two-hadron production in electron-positron
annihilation via one photon up to and including order $1/Q$, where the scale
$Q$ is defined by the (timelike) photon momentum $q$ (with $Q^2 \equiv q^2$)
and given by $Q$ = $\sqrt{s}$. The quantity $Q$ is required to be much larger
than characteristic hadronic scales, a requirement satisfied by the BELLE
experiment for which $\sqrt s \sim 10.5$ GeV. Although this energy is well
below the $Z$-boson mass, we will now consider $\gamma$-$Z$ interference
effects, which could lead to percent level contributions (${\cal
  O}(s/M_Z^2)$).
 
Only leading order $(1/Q)^0$ effects are discussed, since the combination of 
power corrections of order $1/Q$ and $\gamma$-$Z$ interference effects is 
expected to be negligible (permille level). Here we will only focus on tree 
level, i.e., order $(\alpha_s)^0$. This may be improved upon at a
later stage, following e.g.\ Ref.\ \cite{Olsen:1980cw}. 

To leading order 
the expression for the hadron tensor, including quarks and antiquarks, is
\beq
{\cal W}^{\mu\nu}=3 \int dp^- dk^+ d^2\bm{p}_T^{} d^2 
\bm{k}_T^{}\, \delta^2(\bm{p}_T^{}+
\bm{k}_T^{}-\bm{q}_T^{})\, 
\left. \text{Tr}\left( \overline \Delta (p) 
V^\mu \Delta 
(k) V^\nu \right)\right|_{p^+,k^-} + \left(\begin{array}{c} 
q\leftrightarrow -q \\ \mu \leftrightarrow \nu
\end{array} \right),
\eeq
where for a photon $V^\mu = e \gamma^\mu$ and for a $Z$ boson $V^\mu = g_V
\gamma^\mu + g_A \gamma_5 \gamma^\mu$. We have omitted flavor indices and 
summation. The vector and axial-vector couplings to the $Z$ boson are given 
by:
\begin{eqnarray} 
g_V^j &=& T_3^j - 2 \, Q^j\,\sin^2 \theta_W,\\
g_A^j &=& T_3^j,
\end{eqnarray} 
where $Q^j$ denotes the charge and $T_3^j$ the weak isospin of 
particle $j$ (i.e., $T_3^j=+1/2$ for $j=u$ and $T_3^j=-1/2$ for
$j=e^-,d,s$). 

The lepton tensor is given by (neglecting lepton masses and
polarization) 
\begin{equation} 
L_{\mu\nu}^{ij} (l, l^\prime)
=  C^{ij}\, \left[ 2 l_\mu l^\prime_\nu
+ 2 l_\nu l^\prime_\mu - Q^2 g_{\mu\nu} \right]+ D^{ij} \, 
2i \,\epsilon_{\mu\nu\rho\sigma} l^\rho l^{\prime \sigma} \;,
\label{leptten3}
\end{equation}
where we have defined 
\ba
&& C^{\g\g}=1, \quad C^{\g Z}=C^{Z\g}=e^l g_V^l, 
\quad C^{ZZ}= g_V^l{}^2 + g_A^l{}^2 \;,\\
&& D^{\g\g}=0, \quad D^{\g Z}=D^{Z\g}=-e^l g_A^l, 
\quad D^{ZZ}=- 2 g_V^l g_A^l \;,
\ea 
where $e^l$ denotes the coupling of the photon to the leptons in units of 
the positron charge; $g_V^l$, $g_A^l$ denote the vector and axial-vector
couplings of the $Z$ boson to the leptons, respectively.

The cross section is obtained from the hadron tensor after contraction with
the lepton tensor, here given in
the perpendicular basis, 
\begin{eqnarray}
L^{\mu \nu}_{ij} & = & C^{ij} \,Q^2\, \Biggl[
- \left( 1 - 2y + 2y^2 \right) g_\perp^{\mu \nu}
+ 4y(1-y) \hat z^\mu \hat z^\nu
\nonumber \\ && \qquad \qquad \qquad 
-4y(1-y)\left( \hat l_\perp^\mu\hat l_\perp^\nu +\frac{1}{2}\,g_\perp^{\mu \nu}
\right)
- 2(1 - 2y)\sqrt{y(1-y)}\,\,\hat z_\ph^{\{ \mu}\hat l_\perp^{\nu \}} \Biggr]
\nonumber \\ &&\quad + D^{ij} \,Q^2\, \Biggl[
i\,(1-2y)\,\epsilon_\perp^{\mu \nu}
- 2i \,\sqrt{y(1-y)}\,\,\hat l{}_{\perp\rho} 
\epsilon_\perp^{\rho\,[ \mu}
\hat z_\ph^{\nu ]} \Biggr].
\end{eqnarray}
Here only unsuppressed results arise when both indices $\mu, \nu$ are in the
perpendicular directions. 

\begin{table}[htb]
\caption{\label{EWcontractions}
Contractions of the lepton tensor $L_{\mu \nu}^{ij}$ with tensor
structures appearing in the hadron tensor.}
\begin{center}
\begin{tabular}{ccl}
\\
\hline
\\[-3 mm]
$w^{\mu \nu}$ & {} \hspace{3 cm} {} & $L_{\mu \nu}^{ij} w^{\mu \nu}/(4 Q^2)$ 
\\[2mm] \hline \\[-4 mm] \hline
\\[-2 mm]
$-g_\perp^{\mu \nu}$ & &
$ C^{ij} \, \left( \frac{1}{2} - y + y^2 \right)$ \\[2mm] 
$a_\perp^{\,\{ \mu} b_\perp^{\nu\}} -
(a_\perp \cdot b_\perp)\,g_\perp^{\mu\nu}$ & &
$- C^{ij} \, y \left( 1 - y \right)
\vert \bm{a}_\perp \vert \, \vert \bm{b}_\perp \vert\,\cos(\phi_a+\phi_b)$
\\[2mm] 
$\frac{1}{2} \left(
a_\perp^{\{\mu}\,\epsilon_\perp^{\nu \} \rho} b_{\perp\rho}
+ b_\perp^{\{\mu}\,\epsilon_\perp^{\nu \} \rho} a_{\perp\rho}\right)$ & &
$ C^{ij} \, y\left( 1 - y \right)
\vert \bm{a}_\perp \vert \, \vert \bm{b}_\perp \vert\,\sin(\phi_a+\phi_b)$
\\[2mm] 
= $a_\perp^{\{\mu}\,\epsilon_\perp^{\nu\} \rho} b_{\perp\rho}
-(\epsilon_\perp^{\rho\sigma}a_{\perp \rho} b_{\perp\sigma})\,g_\perp^{\mu\nu}$
& & \\[2mm]
$i\,\epsilon_\perp^{\mu \nu}$ & &
$-D^{ij}\,\,\left(\frac{1}{2}-y\right)$ \\[2mm]
$i\,a_\perp^{\,[ \mu} b_\perp^{\nu ]}$ & &
$-D^{ij}\,\,\left(\frac{1}{2}-y\right)\,
\vert \bm{a}_\perp \vert \, \vert \bm{b}_\perp \vert\,\sin (\phi_b - 
\phi_a)$ \\[2mm]
\hline
\end{tabular}
\end{center}
\end{table}

Using Table~\ref{EWcontractions} one finds for the leading order 
unpolarized cross section, taking into
account both photon and $Z$-boson contributions \cite{Boer01}, 
\ba
\lefteqn{
\frac{d\sigma (e^+e^-\to h_1h_2X)}{dz_1 dz_2 d\Omega 
d^2{\bm q_T^{}}} =
\frac{3\alpha^2}{Q^2}\;z_1^2z_2^2\;\sum_{a,\bar a} \;\Bigg\{ 
          K_1^a(y)\;{\cal F}\left[D_1\overline D_1\right]}
\nonumber\\ && 
       \qquad + \left[K_3^a(y)\cos(2\phi_1)+K_4^a(y)\sin(2\phi_1)\right]\;
             {\cal F}\left[\left(2\,\bm{\hat h}\!\cdot \!
\bm p_T^{}\,\,\bm{\hat h}\!\cdot \! \bm k_T^{}\,
                    -\,\bm p_T^{}\!\cdot \! \bm k_T^{}\,\right)
                    \frac{H_1^{\perp}\overline H_1^{\perp}}{M_1M_2}\right]
\Bigg\}.
\label{EWxs}
\ea
The functions $K_i^a(y)$ (as before, $a$ is the flavor 
index) are defined as 
\ba
K_1^a(y) &=& A(y)\;
\left[ e_a^2+ 2 e_l g_V^l e_a g_V^a \chi_1 + c_1^l c_1^a \chi_2
\right] 
- \frac{C(y)}{2} \; \left[ 2 e_l g_A^l e_a g_A^a \chi_1 + c_3^l c_3^a \chi_2 
\right],\\
K_3^a(y)&=& B(y)\; \left[ e_a^2+ 2 e_l g_V^l e_a g_V^a \chi_1 + c_1^l c_2^a 
\chi_2 \right], \\[1 mm] 
K_4^a(y)&=& B(y)\; \left[ 2 e_l g_V^l e_a g_A^a \chi_3 \right],
\ea
which contain the combinations of the couplings
\begin{eqnarray} 
c_1^j &=&\left(g_V^j{}^2 + g_A^j{}^2 \right),\\
c_2^j &=&\left(g_V^j{}^2 - g_A^j{}^2 \right),\\
c_3^j &=&2 g_V^j g_A^j.
\end{eqnarray}
The propagator factors are given by 
\ba
\chi_1 &=& \frac{1}{\sin^2 (2 \theta_W)} \, \frac{Q^2
(Q^2-M_Z^2)}{(Q^2-M_Z^2)^2 + \Gamma_Z^2 M_Z^2},\\
\chi_2 &=& \frac{1}{\sin^2 (2 \theta_W)} \, \frac{Q^2}{Q^2-M_Z^2} \chi_1,\\
\chi_3 &=& \frac{-\Gamma_Z M_Z}{Q^2-M_Z^2} \chi_1.
\ea
Using $M_Z = 91.1876 \pm 0.0021$ GeV $\approx 91.19$ GeV 
and $\Gamma_Z = 2.4952 \pm 0.0023$ GeV $\approx 2.50$ GeV, and using $\sin^2
\theta_W (M_Z)|_{\overline{MS}} = 0.23113 (15) \approx 0.231$ (and assuming
very slow running of this quantity \cite{Erler:2004in}), we obtain for $Q = 10.50$ GeV 
\beq
\chi_1 \approx 0.019, \quad \chi_2 \approx 0.00036,\quad
\chi_3 \approx - 0.00052.
\eeq
This shows that we can neglect the $\sin 2\phi_1$ asymmetry in Eq.\
(\ref{EWxs}) and also the $ZZ$ contributions in $K_1$ and $K_3$.   
This leads to the following approximations for $K_i^a$ in the lepton center of
mass frame:
\ba
K_1^u(y) & \approx & \frac{e_u^2}{4}\;
\left[ 1.0005 (1+\cos^2 \theta_2) + 0.03 \cos \theta_2 \right],\\
K_1^d(y) & \approx & \frac{e_d^2}{4}\;
\left[ 1.002(1+\cos^2 \theta_2) + 0.06 \cos \theta_2 \right],\\
K_3^u(y)&\approx& \frac{e_u^2}{4}\;\left[ 1.0004 \sin^2 \theta_2\right], \\ 
K_3^d(y)&\approx& \frac{e_d^2}{4}\;\left[ 1.001 \sin^2 \theta_2 \right], \\ 
K_4^{u}(y)& \approx & \frac{e_u^2}{4}\;\left[ 0.00003 \sin^2 \theta_2\right]
\approx 0,\\
K_4^{d}(y)& \approx & \frac{e_d^2}{4}\;\left[ 0.00006 \sin^2 \theta_2\right]
\approx 0.
\ea
In general, the $\gamma$-$Z$ interference corrections are small (permille
level), except for the forward-backward asymmetry ($\sim \cos \theta_2$), 
which is on the few percent level. The latter agrees with the results of Ref.\
\cite{Stuart:ng}, taking into account that their 
angle $\theta$ is defined as the angle between the incoming electron and the 
outgoing 
quark, leading to an additional minus sign for the asymmetry. 

In order to estimate the size of the hadron level forward-backward asymmetry,
we will again consider the case of $\pi^+$ and $\pi^-$ production from only $u$ 
and $d$ quarks. Furthermore, we assume 
$D_1^{u\to\pi^+}(z)=D_1^{\bar d\to\pi^+}(z)$, 
$D_1^{d\to\pi^-}(z)=D_1^{\bar u\to\pi^-}(z)$ and neglect unfavored
fragmentation functions like $D^{d\to\pi^+}(z)$, etc. 
If we define 
\beq
d\sigma \sim \left[ 1+\cos^2 \theta_2 + A_{FB}\cos \theta_2 +
\ldots \right],
\eeq
then these assumptions lead to $A_{FB} \approx $ 4\%.

In conclusion, $\gamma$-$Z$ interference does not lead to any significant 
additional $\phi$ dependence at $\sqrt s \sim 10.5$ GeV. 
It only modifies the $\theta$ distribution with 
a forward-backward asymmetry term of a few percent.  

\section{\label{sec:Jet}Jet frame asymmetry}

In an unpublished study \cite{Delphi} a transverse spin correlation 
similar to the $\cos
2\phi$ in back-to-back jets was experimentally investigated to some
extent using LEP's DELPHI data. 
The following angular dependence of the differential cross
section for correlated hadron production in opposite jets was studied:
\beq
\frac{d\sigma}{d\cos\theta\, d\phi \, d\phi'} \propto 1 + \cos^2 \theta +
c_{TT} S \sin^2\theta\cos(\phi + \phi'). 
\label{Scos}
\eeq
Here $c_{TT}=(|v_q|^2-|a_q|^2)/(|v_q|^2+|a_q|^2)$, with $v_q$
and $a_q$ the vector and axial-vector couplings of quarks to the $Z$
boson, respectively. So in our notation $c_{TT} = c_2^a/c_1^a$. 
$S$ is the analyzing power of the asymmetry to be determined. The azimuthal 
angles $\phi$ and $\phi'$ are of the leading particles in
the two jets with respect to the $q\bar q$ axis in the lepton pair center of
mass frame. 

Such a $\cos(\phi + \phi')$
asymmetry differs from the $\cos 2 \phi_1$ asymmetry discussed thus
far in that now {\em three} momenta in the
final state need to be determined, namely besides two hadron momenta also the
jet axis. Hence there are two azimuthal
angles, $\phi$ and $\phi^\prime$, that need to be measured. Furthermore, as we
will show, the analyzing power $S$ is a different expression of the Collins 
functions. We find that it involves moments of the
functions $H_1^\perp$ and $\overline H{}_1^\perp$, different from the ones in
the correlation Eq.\ (\ref{wXsection-cos2phi_O}) that we considered thus far.

Unlike Ref.\ \cite{Delphi} we will not only include the $Z$ boson, but
also the photon contributions, including photon-$Z$ interference terms 
(except for the interference term analogous to $K_4^a(y) \sin 2\phi_1$, 
which was found to be very small).  

To derive a tree level expression for the analyzing
power $S$ in terms of fragmentation functions, 
we start with the hadron tensor expressed in the transverse basis
\ba
\lefteqn{
{\cal W}_{ij}^{\mu\nu}=12 \, z_1z_2
\int d^2\bm k_T^{}\; d^2\bm p_T^{}\;
\delta^2 (\bm p_T^{}+\bm k_T^{}
         -\bm q_T^{})\Bigg\{ 
       - \bigg[g_T^{\mu\nu} \, C_{ij}^a - i\epsilon_T^{\mu\nu}\, D_{ij}^a 
           \bigg]   D_1\overline D_1}\nonumber\\[2 mm]
&& \hspace{7cm} -\frac{k_T^{\{\mu}p_T^{\nu\}}
              +g_T^{\mu\nu}\,\bm k_T\!\cdot\!\bm p_T\,}{M_1M_2} 
         E_{ij}^a\, H_1^{\perp}\overline H_1^{\perp}
         \Bigg\},
\ea
where in analogy to the lepton tensor Eq.\ (\ref{leptten3}) we have defined
\ba
&& C_{\g\g}^a=e_a^2, \quad C_{\g Z}^a=C_{Z\g}^a=e^a g_V^a ,
\quad C_{Z Z}^a = c_1^a\;,\\ 
&& D_{\g\g}^a=0, \quad D_{\g Z}^a=D_{Z\g}^a=-e^a g_A^a,
\quad D_{Z Z}^a = -c_3^a\; , \\
&& E_{\g\g}^a=e_a^2, \quad E_{\g Z}^a=E_{Z\g}^a=e^a g_V^a ,
\quad E_{Z Z}^a = c_2^a\;.
\ea 

In previous sections we have transformed such hadron tensor expressions into
the perpendicular basis of the GJ frame, after which the contraction with the
lepton tensor yields the cross section. Now we will follow Ref.\
\cite{Delphi}. If one would determine the jet-axis, which is identified with
the $q\bar q$ axis, or as an approximation to it the thrust-axis, 
then a measurement of the transverse momenta of the
leading particles in the two jets compared to the jet momentum is a
determination of the transverse momenta of the quarks compared to the leading
hadrons they fragment into. One can then keep the cross section differential
in the azimuthal angles of the transverse momentum of the quarks, after which
the $\bm q_T^{}$ integration can be safely done (as opposed to the case of the
$\cos 2\phi_1$ asymmetry) and it will not average to zero unless one integrates
over the azimuthal angles. In this way one will arrive at an expression
involving the moments 
\beq
F^{[n]}(z_i) \equiv \int d |\bm{k}_{T}|^2 \;
\left[\frac{|\bm{k}_{T}|}{M_i}\right]^n\;  
F(z_i,|\bm{k}_{T}|^2), 
\eeq
for $n=0$ and $n=1$. The latter is often referred to as the ``half-moment'' and
also written as $F^{(1/2)}$. 

To make the
transformation from the frame in which $P_1$ and $P_2$ are
collinear (for which we employed the transverse basis) 
into the lepton pair center of mass frame where the $q \bar q$
axis defines the $\hat{\bm{z}}$ axis, which we will refer to as the jet frame, 
a new perpendicular basis will be defined (the new perpendicular directions 
will also be indicated by $\perp$). We will choose:
\begin{eqnarray}
\hat t^\mu& \equiv & \frac{q^\mu}{Q},\\
\hat z^\mu & \equiv & \frac{k^\mu-p^\mu}{Q}. 
\end{eqnarray}
We find that
\beq 
g_\perp^{\mu\nu} = g_T^{\mu\nu} -
\frac{\sqrt{2}}{Q} \left(p_T^{\{\mu} n_+^{\nu\} }+ k_T^{\{\mu} n_-^{\nu\} }
\right) - \frac{2}{Q^2} p_T^{\{\mu} k_T^{\nu\} }.
\eeq
Hence, $P_{1\perp}= -z_1\, k_T$ and $P_{2\perp}= -z_2\, p_T$, up to $1/Q^2$ 
corrections. 
 
Since the $\hat z$ axis is defined differently now, the contraction of the
hadron tensor with the
lepton tensor yields somewhat different expressions. 
We give the relevant 
contractions in Table \ref{contractions2}. 
The functions $A,B$ and $C$ are the same functions of
$y=P_2\cdot l/ P_2\cdot q$ as before. If one expresses the lepton momentum $l$
in terms of $\hat t$ and $\hat z$, i.e., $l=Q\, \hat t /2 + \sqrt{Q^2/4 -
\bm{l}^2_\perp} \, \hat z + l_\perp$, one finds that $A(y) =1/2 - B(y)$ and
$C(y)=-\sqrt{1-4\, B(y)}$, with $B(y)=\bm{l}^2_\perp/Q^2$.  
\begin{table}[htb]
\caption{\label{contractions2}
Contractions of the lepton tensor $L_{\mu \nu}$ with tensor
structures appearing in the hadron tensor, for the jet frame.}
\begin{center}
\begin{tabular}{cll}
\\
\hline
\\[-3 mm]
$w^{\mu \nu}$ & \mbox{} \qquad \qquad \mbox{} 
& $L_{\mu \nu} w^{\mu \nu}/(4 Q^2)$ 
\\[2 mm] \hline \\[-4 mm] \hline
\\[-2 mm]
$-g_T^{\mu \nu}$ & &
$ c_1^l\,A(y)$ \\[2 mm]
$k_T^{\,\{ \mu} p_T^{\nu\}} +
(\bm{k}_T \cdot \bm{p}_T)\,g_T^{\mu\nu}$ & & 
$- c_1^l \, B(y) \,
\vert \bm{P}_{1\perp} \vert \, \vert \bm{P}_{2\perp} 
\vert\,\cos(\phi+\phi')/(z_1 z_2)$ \\[2 mm]
$i\,\epsilon_T^{\mu \nu}$ & & 
$c_3^l\,C(y)/2$ \\[2 mm]
\hline
\end{tabular}
\end{center}
\end{table}

Using Table \ref{contractions2} we obtain in leading order in $1/Q$ and
$\alpha_s$ the following expression for the cross section differential in
$\phi$ and $\phi'$, in case of unpolarized final state hadrons:
\begin{eqnarray}
\frac{d\sigma (e^+e^-\to h_1h_2X)}{dz_1 dz_2 d\Omega 
d\phi \, d\phi'} & = & 
\sum_{a,\overline a}\;\frac{3\,\alpha^2}{Q^2}
\;z_1^2z_2^2\;\Bigg\{K_1^a(y)\; D_1^{[0]a}(z_1)\overline D{}_1^{[0]a}(z_2)
\nn\\
&& + K_3^a(y)\;\cos(\phi+\phi')\,
H_1^{\perp [1] a}(z_1)\overline H{}_1^{\perp [1] a}(z_2)
\Bigg\}.
\label{phiphiprime}
\end{eqnarray}
This is to be compared with Eq.\ (\ref{LO-OOO}). For simplicity of 
comparison we will now consider the one-flavor case ($u$-quark dominance). 
Hence, we find for the analyzing power $S$ of Eq.\ (\ref{Scos}) 
\beq
S = \frac{H_1^{\perp [1]}(z_1)\overline H{}_1^{\perp [1] }(z_2)}{
D_1^{[0]}(z_1) \overline D{}_1^{[0]}(z_2)}.
\eeq
This can be compared to the expression for the weighted $\cos 2 \phi_1$
asymmetry (cf.\ Eq.\ (\ref{ratio})):
\begin{equation}
\frac{\left< \frac{Q_T^2}{4M_1M_2}\,\cos 2\phi_1  \right>}{\left< 1 \right>}
= \frac{K_3(y)}{K_1(y)} S'
\label{ratiomod}
\end{equation}
where 
\beq
S' = \frac{H_1^{\perp(1)}(z_1)\overline
H{}_1^{\perp(1)}(z_2)}{D_1(z_1)\overline D{}_1(z_2)}.
\eeq

If one now simply assumes a Gaussian $\bkt$-dependence of the functions, i.e., 
\beq
H_1^\perp(z, \bm{k}_T^\prime{}^2)= H_1^\perp(z)\, R^2\, \exp(-R^2\,
\bkt^2)/(\pi \, z^2)
\eeq 
and similarly for $D_1(z, \bm{k}_T^\prime{}^2)$, and equal masses and widths, 
then one finds
that 
\beq
S= \frac{\pi}{4\, R^2\, M^2}\, 
\frac{H_1^\perp(z_1)\overline H{}_1^\perp(z_2)}{D_1(z_1)\overline D{}_1(z_2)}
\eeq  
and 
\beq
S' = \frac{1}{4\,R^4\, M^4}
\, \frac{H_1^{\perp}(z_1)\overline H{}_1^{\perp}(z_2)}{D_1(z_1)\overline 
D{}_1(z_2)} 
\eeq
Note that in these expressions only the Gaussian width of $H_1^\perp$
appears. 
Therefore, one finds:
\beq 
S'= \frac{1}{\pi\,R^2\, M^2} S.
\eeq
For a pion we assume $R\, M \approx 0.5$, such 
that $S' \approx 4S/\pi$. As $S'$ was expected to be on the few percent level,
we also expect the analyzing power of the $\cos(\phi + \phi')$
asymmetry in the jet frame to be of that magnitude. 

If one considers Gaussian transverse momentum dependence one does not need to
resort to weighting to deal with the convolutions. One can consider 
simply the $Q_T^2$-integrated $\cos 2 \phi_1$ asymmetry 
(numerator and denominator integrated separately). 
One arrives at:
\ba
S'' & \equiv & \frac{\int d \bm{q}_T^{2}{\cal F}\left[\left(2\,\hat{\bm{h}}\!\cdot \!
\bm k_T^{}\,\,\hat{\bm{h}}\!\cdot \!
\bm p_T^{}\,
                    -\,\bm k_T^{}\!\cdot \!
\bm p_T^{}\,\right) \; H_1^{\perp}\overline H_1^{\perp}
\right]}{M^2 \int d \bm{q}_T^{2} {\cal F}\left[D_1\overline D_1\right]}\nn \\[2 mm]
& = & \frac{1}{2\,R^2\, M^2}
\, \frac{H_1^{\perp}(z_1)\overline H{}_1^{\perp}(z_2)}{D_1(z_1)\overline 
D{}_1(z_2)} = \frac{2}{\pi} S,
\ea
which is somewhat smaller than the analyzing power $S$ of the
$\cos(\phi+\phi')$ asymmetry. All this indicates that changing from the GJ
frame to the jet frame does not modify the analyzing power of the Collins
effect asymmetry much. 

\section{\label{sec:Scale}Scale dependence of the Collins effect asymmetry} 

Thus far we have only discussed the azimuthal dependences that arise at tree
level. This means that the expressions are valid in the region where the
observed transverse momentum $Q_T$ is small compared to the hard scale $Q$,
applicable only in the region of intrinsic transverse momentum. After the
extraction of the Collins function at BELLE, which is performed at one
particular $Q^2$ value, the extracted functions will be used in asymmetries at
different energies, for instance at lower $Q^2$ for SIDIS or higher $Q^2$ for
a comparison to LEP1 data. For this one would need to know the scale
dependence of the Collins function and asymmetry. However, the study of the
scale dependence of expressions involving TMDs is not far developed.  Here we
will follow (and improve) the discussion of this issue given in Ref.\
\cite{Boer01}.

It is clear that collinear factorization is not the right framework to address
the question of scale dependence of the Collins effect asymmetry. Rather, for
small $Q_T$ the so-called Collins-Soper (CS) factorization theorem 
\cite{CS81} is of
relevance (see also Ref.\ \cite{Anikin:2006hm} for a discussion of
factorization and transverse momentum in two-hadron inclusive $e^+ e^-$
annihilation). It applies to the differential cross section:
\beq
\frac{d\sigma}{dz_1
dz_2 d\Omega {d^2 \bm{q}_T^{}}}
\eeq 
Collins \& Soper \cite{CS81} proved their factorization theorem for
the process of interest here: $e^+e^- \to h_1 \, h_2 \, X$, but without 
inclusion of quark spin effects. A similar factorization
for SIDIS and Drell-Yan, including polarization, was discussed more recently 
by Ji, Ma \& Yuan \cite{JMY}, with some small 
differences w.r.t.\ Ref.\ \cite{CS81}. 

In this section we will study the scale dependence of the Collins 
effect asymmetry taking CS factorization as our 
starting point. It allows us to
obtain the dominant $Q^2$ dependence of the asymmetry at small
$Q_T^2$, but the analysis also extends the range of applicability of the
asymmetry results from the region of intrinsic transverse
momentum to the region of moderate $Q_T^{}$ values (still under the 
restriction that $Q_T^2 \ll Q^2$). The consequences discussed
below affect all transverse momentum dependent
azimuthal spin asymmetries. It turns out that such azimuthal
asymmetries generally suffer from Sudakov 
suppression with increasing $Q^2$ in the region where the transverse 
momentum $Q_T$ is much smaller than the large energy scale $Q$. This Sudakov 
suppression stems from {\em soft\/} gluon radiation. In Ref.\ 
\cite{Collins-93b} Collins remarks that Sudakov factors 
will have the effect of diluting the (Collins effect) single spin asymmetry 
in semi-inclusive DIS, due to broadening of the 
transverse momentum distribution by soft gluon emission.
Here we will study this 
effect in a quantitative way for the $\cos 2 \phi $
Collins-effect 
asymmetry in $e^+e^- \to h_1 \, h_2\, X$. It shows that
tree level estimates of such asymmetries tend to yield overestimates and
increasingly so with rising energy. 

\subsection{\label{sec:CS81}Collins-Soper factorization}

In the CS formalism the differential cross section at small
$Q_T^2/Q^2$ is written as
\beq
\frac{d\sigma}{dz_1 dz_2 d\Omega {d^2 \bm{q}_T^{}} } = \int d^2 {b} 
\, e^{-i {\bm{b} \cdot \bm{q}_T^{}}} \tilde{W}({\bm{b}}, Q;
  z_1, z_2) + {\cal O}\left(\frac{Q_T^2}{Q^2}\right), 
\label{CS81xs}
\eeq
where 
\ba
\tilde{W}({\bm{b}}, Q; z_1, z_2) & = &
{\sum_{a}\,\tilde{D}_1^{a}(z_1,{\bm{b}};1/b,\alpha_s(1/b))}
{\sum_{b}\, \tilde{D}_1^{b}(z_2,{\bm{b}};1/b,\alpha_s(1/b))} \nn\\ 
& & \times 
e^{-S(b,Q)} \,H_{ab}\left(Q;\alpha_s(Q)\right)\,
\tilde U(b;1/b,\alpha_s(1/b)). 
\label{CS81W}
\ea
Here $\tilde{D}_1(z,\bm{b})$ is the Fourier transform of the
transverse momentum dependent fragmentation function 
$D_1(z,z^2\bm{k}_T^2)$; $e^{-S(b,Q)}$ is the Sudakov form factor;   
$H$ is the partonic hard scattering part; and, $\tilde U$ is called the 
soft factor. The fragmentation functions and
the soft factor are taken at the scale $\mu=1/b$.
Furthermore, note that there are no integrals over momentum fractions, 
those appear in the large $Q_T$ or equivalently, small-$b$ limit only. 
Hence, the $z_i$ in Eq.\ (\ref{CS81W}) 
are the observed momentum fractions.  

The Sudakov form factor arises due to
summation of soft gluon contributions. This is in contrast to more
inclusive cross sections for which there is often a cancellation of
such soft gluon contributions. 
At values $b^2=\bm{b}^2 \ll 1/\Lambda^2$, the Sudakov form factor is
perturbatively calculable and of the form 
\beq
S(b,Q)=\int_{b_0^2/b^2}^{Q^2} \frac{d \mu^2}{\mu^2} \left[ 
A(\alpha_s (\mu)) \, \ln \frac{Q^2}{{\mu}^2} + 
B (\alpha_s (\mu)) \right],
\label{sud}
\eeq
where $b_0=2\exp(-\gamma_E) \approx 1.123$ (here we use the usual CS
factorization constants $C_1=b_0, C_2=1$). 
One can expand the functions $A$ and $B$ in $\alpha_s/\pi$ and the first few 
coefficients are known, cf.\ e.g.\ \cite{Davies-Stirling,Weber}. At the
leading logarithm (LL) level one needs to take into account only the 
first term in the expansion of $A$: $A^{(1)} =  C_F/\pi$, which is the same
for unpolarized and polarized scattering. It leads to the 
double leading logarithmic
approximation (DLLA) result \cite{ParisiPetronzio}:
\beq
S(b,Q)=C_F \frac{\alpha_s(Q)}{2\pi} \log^2 \left(b^2 Q^2\right).
\label{SDLA}
\eeq
Including the running of $\alpha_s$ leads to the
expression \cite{Frixione}
\beq
S(b,Q)=-\frac{16}{33-2n_f} \left[ \log\left(\frac{b^2 Q^2}{b_0^2}\right)+
\log\left(\frac{Q^2}{\Lambda^2}\right)\; \log\left[1- \frac{\log\left(b^2 
Q^2/b_0^2\right)}{\log\left(Q^2/\Lambda^2\right)} \right]\right].
\label{sudactual}
\eeq
We will take for the number of flavors
$n_f=5$, since we are
interested in energies just below the $\Upsilon(4S)$, which is above
the $b \bar{b}$ threshold. Furthermore, we take 
$\Lambda_{QCD}=200 \, \text{MeV}$. 

Using only the perturbative expression for the Sudakov factor in the 
cross section expression (\ref{CS81xs}) is valid for $Q^2$ 
very large, when the restriction $b^2 \ll 1/\Lambda^2$ is justified. 
If also 
$b^2 \simorder 1/\Lambda^2$ contributions are important ($\mu^2
\simorderr \Lambda^2$), for example at small $Q_T$, then 
one needs to include a nonperturbative Sudakov factor. This can be
done for instance via the introduction of a $b$-regulator 
\cite{CS85e}\footnote{An alternative method has been put forward in 
Ref.\ \cite{Laenen:2000de}.}: 
$b \to b_*=b/\sqrt{1+b^2/b_{\max}^2}$, such that $b_*$ stays always smaller
than $b_{\max}$. 
Usually $b_{\max}=0.5 \; \text{GeV}^{-1}$, such that 
$\alpha_s(b_0/b_* \simorder 2) \simorderr 0.3$. For the function 
$\tilde{W}(b_*)$ a perturbative expression for the Sudakov factor can thus 
be used. 

One can rewrite $\tilde{W}(b)$ as:
\beq
\tilde{W}(b) \equiv \tilde{W}(b_*) \, e^{-S_{NP}(b)} .
\eeq
In general the nonperturbative Sudakov factor is of the form \cite{CS85e}
\beq
S_{NP}(b,Q/Q_0) = g_1(z_1,b) + 
g_2(z_2,b) + \ln(Q^2/Q_0^2) g_3(b), 
\label{genNPform}
\eeq
where $Q_0 \approx 1/b_{\max}$ is the lowest scale for
which one trusts perturbation theory. 
The $g_{i}$ functions ($i=1,2,3$) are not calculable in perturbation theory 
and need to be fitted to data. In fact, for the Drell-Yan process 
they have been shown to be necessary to include in order to describe available
data \cite{Landry}. Note that $S_{NP}$ is in general $Q^2$ dependent, unlike
what one may at first thought expect for a nonperturbative quantity. 

Due to the lack of knowledge of the nonperturbative Sudakov factor for
the case of interest here, the quantitative results obtained below about the
size and $Q^2$ dependence of the suppression should not be taken too
literally. Nevertheless, one can draw conclusions about what determines
the $Q^2$ dependence of the transverse momentum distribution 
of the asymmetries and about the size of the suppression for generic 
nonperturbative Sudakov factors. 

Concerning the other factors in the CS factorization expression, it
should be noted that the one-loop expression for $H$ and $\tilde U$ were
obtained in Ref.\ \cite{CS81}. At the scale $\mu=1/b$ the
$b$-dependence of $\tilde U$ disappears at order $\alpha_s$ in the
$\overline{MS}$ scheme. 

Due to the integration over $b$, the CS factorization expression explicitly
requires knowledge of the scale dependence of the $b$-dependent
fragmentation functions. In Ref.\ \cite{Boer01} this dependence was
assumed to be weak (logarithmic) and neglected, but here we will include
it and use that it is determined by the quark field renormalization. 
As already mentioned by 
Collins \& Soper \cite{CS81} in some cases it may be more convenient to take 
instead of the varying scale $\mu=1/b$, a fixed scale $M_0 \simorder 1$ GeV. 
That scale was called 
$\mu_L$ by Ji {\em et al.} \cite{JMY,Idilbi} and here we will call it $Q_0$. 
Using the
renormalization group equations given in Ref.\ \cite{CS81}, the CS 
factorization expression (\ref{CS81W}) can be rewritten as:
\ba
\tilde{W}(\bm{b}, Q; z_1, z_2) & = &
\sum_{a}\,\tilde{D}_1^{a}(z_1,\bm{b};Q_0,\alpha_s(Q_0))
\sum_{b}\,\tilde{D}_1^{b}(z_2,\bm{b};Q_0,\alpha_s(Q_0)) \nn\\ 
& & \times e^{-S(b,Q,Q_0)}\,H_{ab}\left(Q;\alpha_s(Q)\right)\, 
\tilde{U}(b;Q_0,\alpha_s(Q_0)).
\label{Wtilde2}
\ea
This has the advantage that the 
$b$-dependent fragmentation functions are always considered at 
the same scale $Q_0$ when integrating over $b$.

We first consider the above 
expression in the perturbative regime: $b \leq 1/Q_0$.
The one-loop expressions for $H$, $\tilde U$ and $S$ are 
\ba 
H_{ab}\left(Q;\alpha_s(Q)\right) & \propto & \delta_{b \bar{a}} e_a^2 
\left(1 + \alpha_s(Q^2) F_1 + {\cal O}(\alpha_s^2) \right),\\
\tilde U(b;Q_0,\alpha_s(Q_0)) & = & 1 - \frac{\alpha_s(Q_0^2)}{\pi}
C_F \left( \log Q_0^2 b^2 + F_2 \right) + {\cal O}(\alpha_s^2), 
\label{leadingtermu}\\
S(b,Q,Q_0) & = & C_F \int_{Q_0^2}^{Q^2} \frac{d \mu^2}{\mu^2}
\frac{\alpha_s(\mu)}{\pi} \; \left[ \log \frac{Q^2}{{\mu}^2} + 
\log Q_0^2 b^2 + F_3 \right]. \label{leadingterms}
\ea
Here $F_i$ denote renormalization-scheme-dependent finite terms which will be 
neglected in the following. 
Following Ref.\ \cite{Idilbi}, one might also consider dropping 
in the expression between brackets in Eq.\ (\ref{leadingterms}) the
second term with respect to the first one in order to arrive
at the double leading log expression. This is because $\log Q_0^2 b^2$ is in
general not a large log, since the small-$b^2$ ($\sim 1/Q^2$) region
contributes little for small $Q_T$. In that approximation $S$ is not a
function of $b$ anymore. It leads to the double log term in 
$S$ $\propto \log^2 Q^2/Q_0^2$. 
This means there is no coupling between $Q^2$ and $Q_T^2$
(ignoring for simplicity the $Q^2$ dependence of $S_{NP}(b,Q/Q_0)$), such that
the Sudakov factor $\exp(-S)$ will cancel in asymmetry ratios. As a
consequence, azimuthal asymmetries will be $Q^2$-independent in this
approximation.

However, the $\log Q_0^2 b^2$ term in $S$ is important to keep in order
  to obtain the leading $Q^2$-dependence of azimuthal
asymmetries.  
If one keeps this term, then one 
obtains (ignoring the running of $\alpha_s$ for the moment):
\beq
S(b,Q,Q_0) = C_F \frac{\alpha_s(Q)}{2\pi} \; 
\left[ \log^2 Q^2 b^2 - \log^2 Q_0^2 b^2 \right].
\eeq
In this expression the second term {\em can\/} safely be
ignored, if one is interested in the leading $Q^2$-dependence of
asymmetries only. 
The average $b^2$ probed is of order $1/Q_T^2$, such that in
the small-$Q_T^2$ region (of order $Q_0^2$): 
$\log^2 Q^2 b^2 \gg \log^2 Q_0^2 b^2$. The result then coincides 
with the well-known DLLA expression (\ref{SDLA}).
Upon dropping all other small logarithms and finite terms, this leads to:
\ba
\tilde{W}(\bm{b}, Q; z_1, z_2) & \propto & 
{\sum_{a}e_a^2 \,\tilde{D}_1^{a}(z_1,\bm{b};Q_0,\alpha_s(Q_0))}
\, \tilde{D}_1^{\bar{a}}(z_2,\bm{b};Q_0,\alpha_s(Q_0)) \nn\\ 
& & \times \exp\left( -C_F \frac{\alpha_s(Q)}{2\pi} \; \log^2 Q^2 b^2 \right). 
\ea
This expression, which we will refer to below as the DLLA of $\tilde{W}$,
should capture the dominant $Q^2$-dependence of the differential 
cross section, for $Q^2 \gg Q_0^2$ and bearing in mind that this
becomes a worse approximation as $Q^2$ increases. In the numerical
results presented below we will include also the subleading 
logarithms explicitly given above and the one-loop running of $\alpha_s$.

\subsection{\label{sec:Num}Numerical study of the $Q^2$ dependence of the Collins
  effect asymmetry} 

In order to study the expression Eq.\ (\ref{CS81xs}) with $\tilde{W}$ of
Eq.\ (\ref{Wtilde2}), we will assume again a 
Gaussian transverse momentum dependence for $D_1(z, z^2\bm{k}_T^2)$:
\beq
D_1(z, z^2\bm{k}_T^2) = D_1(z) \; R_u^2 
\exp(-R_u^2 \bm{k}_T^2) / \pi z^2 \equiv 
D_1(z) \; {\cal G}(|\bm{k}_T^{} |; R_u)/ z^2, 
\label{Gauss}
\eeq
such that the Fourier transform is
\beq
\tilde{D}{}_{1}^{}(z,b^2) 
= D_{1}(z) \; \exp\left(-\frac{b^2}{4 R_u^2} \right)/z^2.
\label{GaussFT}
\eeq  
We do not need to worry about any scale dependence of the width $R_u$, 
since by construction
the fragmentation functions are always at scale $Q_0$. Here we will not aim
to connect to the large-$Q_T$ behavior of the cross section, where the
Gaussian behavior is certainly not correct anymore. This fact and the 
assumed factorization of $b$ and $z$ dependence implies however that 
it is not guaranteed that the function $D_1(z)$ that appears here is indeed
exactly equal to the integrated fragmentation function $D_1(z;Q_0^2)$
at the scale $Q_0^2$. Therefore, the safest thing may be to construct ratios of
asymmetries where the functions $D_1(z)$ drop out, cf.\ Sec.\ \ref{sec:Ratios}. 
After these words
of caution we will make some further pragmatic assumptions in order to
be able to numerically investigate the $Q^2$ dependence of the Collins
effect asymmetry. 

The $b$-dependent piece of the fragmentation function $\tilde{D}_1(z,b^2)$,
taken to be a Gaussian here, can be viewed as the $Q^2$-independent part of 
$S_{NP}$, because they
are simply indistinguishable (ignoring the subtlety of using $b$ here 
instead of $b_*$). Any remaining $z$ dependence that does not
factorize can in principle also be included in $S_{NP}$. 
We further assume that $S_{NP}$ is flavor and 
spin independent, i.e.\ it is
the same for all fragmentation functions. The assumption of spin
independence will get
better as $Q^2$ becomes larger. Refinements can be included at a later
stage if the accuracy of the data demands it. 
Here we will take for the nonperturbative Sudakov factor 
the parameterization by Ladinsky \& Yuan, which was fitted to the
transverse momentum
distribution of $W/Z$ production in $p p$ $(p \bar{p})$ scattering 
\cite{Ladinsky-Yuan}. We will use it with the additional simplifying choice of 
$x_1 x_2 =10^{-2}$:
\beq
S_{NP}(b,Q/Q_0) = g_1 b^2 + g_2 b^2
\ln\left(\frac{Q}{2Q_0}\right),
\label{LY94}
\eeq
with $g_1=0.11\,\text{GeV}^{2}, g_2=0.58\,\text{GeV}^{2}, 
Q_0=1.6 \,\text{GeV}$ and $b_{\max}= 0.5 \,\text{GeV}^{-1}$. Note that  
for $Q=10$ GeV the $Q^2$-independent part is negligible, which justifies
ignoring the above-mentioned subtlety. 

Although
quantitatively the results will depend considerably on this choice of $S_{NP}$
(cf.\ Ref.\ \cite{Boer01} for a detailed discussion), the $Q^2$
dependence of the asymmetry turns out not to be very sensitive to
it. Nevertheless, 
our results underline the importance of a good determination
of $S_{NP}$ for the proper extraction of the Collins function, when going
beyond tree level, see Sec.\ \ref{sec:SNP}. 

The nonperturbative Sudakov factor allows us to integrate $\tilde{W}$
safely over the large $b$ region, but one also has to deal with the
region of very small $b$. $S(b,Q,Q_0)$ at {\em very\/} small $b^2 <
1/Q^2$ requires regularization in order to ensure the correct limiting
behavior. At small $Q_T$ this issue becomes less important as
$Q^2$ increases. Usually $S$ is regulated by the replacement 
\cite{ParisiPetronzio}:
\beq
\log^2 (Q^2b^2) \longrightarrow \log^2 (Q^2b^2+1).
\label{replace}
\eeq

In summary, we consider the following pragmatic simplifying 
steps: we assume a factorized Gaussian dependence of the fragmentation
functions; we ignore flavor and spin dependence of this Gaussian 
dependence; we
take a generic nonperturbative Sudakov factor known from $pp$
scattering; we do not worry about matching to high $Q_T$; and, 
we drop the finite terms $F_i$. All this leads to a manageable 
expression for $\tilde{W}$, which in DLLA becomes:
\ba
\tilde{W}(\bm{b}, Q; z_1, z_2) & \propto & 
{\sum_{a}e_a^2 \,D_1^{a}(z_1)} \; D_1^{\bar{a}}(z_2) \nn\\ 
& & \times 
\exp\left( -C_F \frac{\alpha_s(Q)}{2\pi} \; \log^2 \left(Q^2
      b_*^2+1 \right) \right) \; \exp\left(-S_{NP}(b,Q/Q_0)\right), 
\ea
But as said, below we will also include the single logarithms and the running
of $\alpha_s$, and doing so leads to qualitatively similar, but 
quantitatively different results compared to earlier results of
\cite{Boer01}, where $\mu=1/b$ was used. 

The final step to be taken is to include the Collins fragmentation
function. This can be done via the replacement of $\tilde{D}(z,b^2)
\to \tilde{\Delta}(z,\bm{b})$, where for unpolarized hadron production:
\beq
\tilde{\Delta}(z,\bm{b}) = \frac{M}{4}\,\Biggl\{
\tilde{D}_1(z,b^2)\, \frac{\slsh{\! P}}{M} + 
\left(\frac{\partial}{\partial b^2} \tilde{H}_1^\perp (z,b^2)
\right) \, \frac{2 \slsh{\bm{b}} \slsh{\! P}}{M^2} \Biggl\},
\eeq
which is the Fourier transform of Eq.\ (\ref{DeltaexpB}) (we have
included a factor $P^-$ into the definition of $\Delta$). Since the 
second term is $\bm{b}$-odd, it leads to a different $Q^2$-dependence 
than the first term. This leads to the $Q^2$ dependence of the
asymmetry. 

A model for the transverse momentum dependence of the 
function $H_1^\perp$ is needed. As for $D_1$ we will simply assume a
Gaussian form: $H_1^{\perp a}(z,z^2\bm{k}_T^2) = H_1^{\perp a}(z) \; 
{\cal G}(|\bm{k}_T^{} |; R)/ z^2$. Strictly speaking, the
radius $R$ should be taken larger than $R_u$ of the unpolarized
function $D_1$, such as to satisfy the bound \cite{Bacchetta:1999kz} 
\beq
|\bm{k}_T^{}| \; \left| H_1^{\perp}(z,|\bm{k}_T^{}|) \right| \leq M_h \; 
D_1(z,|\bm{k}_T^{}|),
\label{boundagain}
\eeq
for all $|\bm{k}_T^{} |$. But since we will include this (in principle spin
dependent) Gaussian dependence into $S_{NP}$, we will approximate $R$
by $R_u$, because the (spin independent) $Q^2$ dependent part of
$S_{NP}$ dominates anyway. Note that equating $R$ by $R_u$ is not
allowed in tree level treatments of the asymmetry, as no fall-off with
$Q_T$ will be obtained otherwise. 

As a further simplification, we will assume that the fragmentation
functions for both hadrons are Gaussians of equal width, i.e.\ we take 
$R_{u1}=R_{u2}=R_u$ and $R_1=R_2=R=R_u$. Also we take $M_1=M_2=M$. All
these simplifications can be easily undone when needed. 

With all these steps in place, we can write the Collins effect
asymmetry in the CS formalism in a form similar to that obtained at
tree level in Sec.\ \ref{sec:Unintegrated}.
If we define the asymmetry $A(\bm{q}_T^{})$ as the analyzing power of
the $\cos 2\phi_1$ asymmetry :
\beq
\frac{d\sigma (e^+e^-\to h_1h_2X)}{dz_1 dz_2 d\Omega d^2{\bm q_T^{}}} \propto 
\left\{ 1 + \cos 2\phi_1 A(\bm{q}_T^{}) \right\},
\eeq
then at tree level we obtained  
\ba
A(\bm{q}_T^{}) & = & \frac{
\sum_{a}\;K_3^a(y)\;{\cal F}\left[\left(2\,\bm{q}_T^{}\!\cdot \!
\bm p_T^{}\,\,\bm{q}_T^{}\!\cdot \! \bm k_T^{}\,
                    -\,\bm{q}_T^{2}\,\bm p_T^{}\!\cdot \! \bm k_T^{}\,\right)
                    H_1^{\perp}\overline H{}_1^{\perp}
\right]}{Q_T^2 {M_1M_2}\sum_{b}\; K_1^b(y) \; 
{\cal F}\left[D_1\overline D_1\right]}.
\label{AQT}
\ea
Here we taken the flavor summation out of the definition of ${\cal
  F}[\ldots]$ and replaced $e_a^2 A(y) \to K_1^a(y)$ and 
$e_a^2 B(y) \to K_3^a(y)$ compared to Sec.\ \ref{sec:Unintegrated}, in
order to include $\gamma$-$Z$ interference effects, although we do ignore
the $\sin 2\phi_1$ Collins effect asymmetry that was estimated to be
small in Sec.\ \ref{sec:Estimate}. We have also multiplied the asymmetry by a 
trivial factor $Q_T^2/Q_T^2$ in order to be able to replace $\bm{\hat
  h} \to \bm{q}_T^{}$. This might seem problematic at $Q_T^2=0$, but
  that turns out not to be a problem
as the asymmetry has a kinematic zero at $Q_T=0$, because $\bm{\hat h}$ 
cannot be defined in that case.

At tree level 
\begin{equation} 
{\cal F}\left[D\overline D\, \right]\equiv \;
\int d^2\bm p_T^{}\; d^2\bm k_T^{}\;
\delta^2 (\bm p_T^{}+\bm k_T^{}-\bm 
q_T^{})  D^a(z_{1},z_{1}^2\bm{p}_T^2) 
\overline D{}^a(z_{2},z_{2}^2\bm{k}_T^2).
\label{conv}
\end{equation}
In order to apply the CS factorization expression we can replace in 
Eq.\ (\ref{conv}) \cite{Boer01}  
\beq
\delta^2(\bm{p}_T^{}+\bm{k}_T^{}-\bm{q}_T^{})\to \int 
\frac{d^2 \bm{b}}{(2\pi)^2} \, e^{-i \bm{b} \cdot
(\bm{p}_T^{}+\bm{k}_T^{}-\bm{q}_T^{})} \, e^{-S} \, \tilde{U},
\label{sudakovreplacement}
\eeq 
leading to (suppressing the flavor indices and the arguments of $S$ and
$\tilde{U}$)   
\ba 
{\cal F}\left[D_1\overline D_1\, \right]& \equiv & \; \int 
\frac{d^2 \bm{b}}{(2\pi)^2} \, e^{i \bm{b} \cdot \bm{q}_T^{}} 
\, e^{-S}\, \tilde{U}\, \tilde{D}_1(z_{1},b^2) \, 
\tilde{\overline D}_1(z_{2},b^2)\nn \\[2 mm]
&=& \frac{1}{2 \pi} \int_0^\infty db \, b \, J_0(b Q_T)\, e^{-S}\, \tilde{U}\,
\tilde{D}(z_{1},b) \, \tilde{\overline D}(z_{2},b).
\label{conv2}
\ea
The numerator in Eq.\ (\ref{AQT}), 
\ba 
\lefteqn{{\cal F}\left[\left(2\,\bm{q}_T^{}\!\cdot \!
\bm p_T^{}\,\,\bm{q}_T^{}\!\cdot \! \bm k_T^{}\,
                    -\,\bm{q}_T^{2}\,\bm p_T^{}\!\cdot \! \bm k_T^{}\,\right)
\, H_1^{\perp}\overline H{}_1^{\perp}\, \right] \equiv  \; \int 
\frac{d^2 \bm{b}}{(2\pi)^2} \, e^{i \bm{b} \cdot \bm{q}_T^{}} 
\, e^{-S}\, \tilde{U}}\nn \\
&& \times \int d^2\bm p_T^{}\; d^2\bm k_T^{}\; \left(2\,\bm{q}_T^{}\!\cdot \!
\bm p_T^{}\,\,\bm{q}_T^{}\!\cdot \! \bm k_T^{}\,
                    -\,\bm{q}_T^{2}\,\bm p_T^{}\!\cdot \! \bm k_T^{}\,\right)
\; e^{-i \bm{b} \cdot (\bm{p}_T^{}+\bm{k}_T^{})}\; 
H_1^{\perp}(z_{1},z_{1}^2\bm{p}_T^2) 
\overline H{}_1^{\perp}(z_{2},z_{2}^2\bm{k}_T^2),
\label{bconv}
\ea
cannot be treated exactly like the denominator. But for Gaussian
transverse momentum dependence of the fragmentation functions the
transverse momentum integrals can be performed. One finds
\ba   
\lefteqn{
\int d^2\bm p_T^{}\; d^2\bm k_T^{}\; \left(2\,\bm{q}_T^{}\!\cdot \!
\bm p_T^{}\,\,\bm{q}_T^{}\!\cdot \! \bm k_T^{}\,
                    -\,\bm{q}_T^{2}\,\bm p_T^{}\!\cdot \! \bm k_T^{}\,\right)
\; e^{-i \bm{b} \cdot (\bm{p}_T^{}+\bm{k}_T^{})}\; {\cal
G}(\bm{p}_T^2;R) \;  
{\cal G}(\bm{k}_T^2;R) 
= } \nn \\
&& \mbox{} \hspace{6 cm} -\frac{1}{4 R^4} \; \left[ 2(\bm{q}_T^{} \cdot \bm{b})^2 - \bm{q}_T^2
\bm{b}^2 \right]\; \exp\left(-\frac{b^2}{2 R^2}\right),
\ea 
which after application to Eq.\ (\ref{bconv}) yields 
\ba
{\cal F}\left[\left(2\,\bm{q}_T^{}\!\cdot \!
\bm p_T^{}\,\,\bm{q}_T^{}\!\cdot \! \bm k_T^{}\,
                    -\,\bm{q}_T^{2}\,\bm p_T^{}\!\cdot \! \bm k_T^{}\,\right)
\; H_1^{\perp}\overline H{}_1^{\perp}\, \right] & = & 
\int 
\frac{d^2 \bm{b}}{(2\pi)^2} \, e^{i \bm{b} \cdot \bm{q}_T^{}} \,
(-\frac{1}{4 R^4}) \; \left[ 2(\bm{q}_T^{} \cdot \bm{b})^2 - \bm{q}_T^2
\bm{b}^2 \right]\nn \\
& & \times e^{-S} \; \tilde{U} \; \tilde{H}_1^\perp (z_{1},b^2) \, 
\tilde{\overline H}_1^\perp (z_{2},b^2). 
\ea
It is important to note that the factor $1/R^4$ stems from 
the Gaussian width of the functions $H_1^\perp(z,\bm{k}_T^{2})$ to be
used in this expression. Therefore, it is not dependent on the scale
$Q$ as discussed above already. 

Putting everything together Eq.\ (\ref{AQT}) can be transformed
into 
\beq
A(Q_T) = \frac{\sum_a \; K_3^a(y) H_1^{\perp a}(z_1) \; 
\overline H{}_1^{\perp a}(z_2)}{4 M^4 R^4
\; \sum_b \; K_1^b(y) D_1^b(z_1) \; \overline D{}_1^b(z_2)} \; 
{\cal A}(Q_T) = \frac{\sum_a K_3^a(y) \; H_1^{\perp
(1) a}(z_1) \; \overline H{}_1^{\perp (1) a}(z_2)}{\sum_b K_1^b(y) \; 
D_1^b(z_1) \; \overline D{}_1^b(z_2)} \; {\cal A}(Q_T), 
\label{kappa3}
\eeq
where 
\beq
{\cal A}(Q_T) \equiv M^2 \, \frac{
\int_0^\infty db \, b^3 \, J_2(b Q_T) \, \tilde{U}(b_*;Q_0,\alpha_s(Q_0)) \,
\exp\left({-S(b_*,Q,Q_0)}\, 
{-S_{NP}(b,Q/Q_0)}\right)}{\int_0^\infty db \, b \, J_0(b Q_T)\, 
\tilde{U}(b_*;Q_0,\alpha_s(Q_0)) \,
\exp\left({-S(b_*,Q,Q_0)}\, {-S_{NP}(b,Q/Q_0)} \right)}.
\label{calAQT}
\eeq
We will employ this expression in combination with Eqs.\ (\ref{leadingtermu}),
(\ref{leadingterms}) and (\ref{LY94}), dropping the finite terms $F_i$, 
but including the one-loop running of $\alpha_s$. As the replacement of 
Eq.\
(\ref{replace}) turns out to have only a minor effect in this expression, in
contrast to the DLLA expression, it will not be included.

\begin{figure}[htb]
\begin{center}
\leavevmode \epsfxsize=10cm \epsfbox{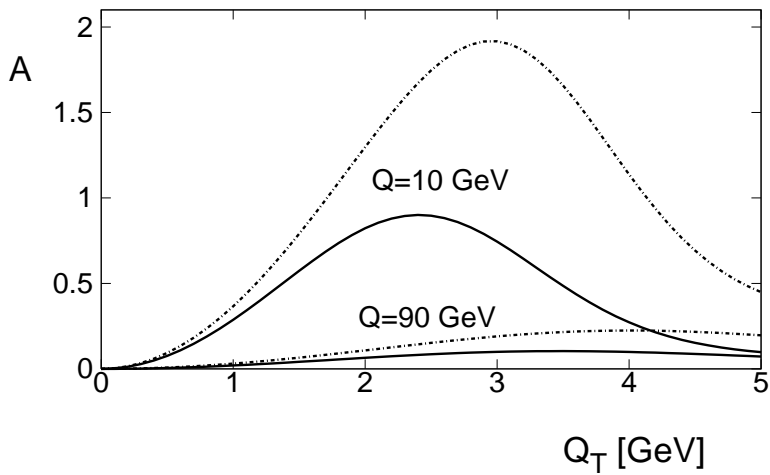}
\caption{\label{4Belle}The asymmetry factor ${\cal A}(Q_T)$ (in units of
$M^2$) at $Q=10 \, \text{GeV}$ 
and $Q = 90 \, \text{GeV}$. The solid curves are obtained with the method 
explained in the text; the dashed-dotted curves are from the earlier analysis
of Ref.\ \cite{Boer01}.}
\end{center}
\vspace{-2 mm}
\end{figure}

In Fig.\ \ref{4Belle} 
the asymmetry factor ${\cal A}(Q_T)$ is given at the scales 
$Q=10 \, \text{GeV}$ and $Q = 90 \, \text{GeV}$, in order to compare
the results for the BELLE and LEP1 scales. 
The solid curves are obtained with the method explained here, whereas 
the dashed-dotted curves are the results obtained from the analysis of
Ref.\ \cite{Boer01}, where $\mu=1/b$ was employed and the scale
dependence of the fragmentation functions was ignored. One observes a 
reduction w.r.t.\ earlier results, but the large Sudakov suppression
with increasing $Q$ remains equally strong, i.e.\ approximately $1/Q$. 
The DLLA result (not displayed) decreases more slowly, but as said it becomes 
a worse approximation as $Q^2$ increases. 

The $\cos 2\phi$ asymmetry has been studied using DELPHI data ($\sqrt{s} =
M_Z$) and the magnitude was found to be small \cite{Efremov}.  This may have
several reasons, one of which could be the Sudakov suppression discussed here.
In that case a small result at LEP1 energy does not imply that also at BELLE
the asymmetry has to be small. In Ref.\ \cite{Efremov:2006qm} a comparison is
made of the Collins functions extracted from the DELPHI and BELLE data (here
it should be emphasized that the DELPHI data analysis remains preliminary and
does not consider possible systematic effects). At higher $Q^2$ the extraction 
of the Collins function from a tree level expression becomes less accurate, as
we will discuss in the next subsection. Therefore, it would be
interesting to also compare the asymmetries directly (rather than the
extracted Collins functions). This could serve as a check of the $Q^2$
dependence of the asymmetry as a whole and thus of the CS formalism.

Of course, one also wants to compare to Collins effect asymmetries
in SIDIS at lower energies (e.g.\ for HERMES $\amp{Q^2}=2.41$ GeV$^2$). 
The question is what to do at smaller values of $Q^2$?
The logarithms that are resummed in the CS formalism are not that large 
to begin with and neither is the relevant $b$-range (set by $Q_0$ and
$Q$). In this case one can set $Q_0=Q$, which means $S(b,Q,Q)=1$ and
$S_{NP}(b)$ $Q^2$-independent, and consider $Q_T \sim M$. 
A reduction to the tree level form occurs up to small 
logarithmic corrections of order $\alpha_s(Q^2) \log Q_T^2/Q^2$. Tree level
analyses should yield reasonable results in this case.

\subsection{\label{sec:Tree}Comparison to tree level}

We will compare the above result for $Q =10\, \text{GeV}$ with the tree
level result (cf.\ Eq.\ (\ref{LO-OOOC})). In the tree level 
expression for the asymmetry it is important to keep Gaussians in 
numerator and denominator different, in order to ensure the bound given 
in Eq.\ (\ref{boundagain}) is satisfied and an asymmetry is obtained 
that falls off at larger $Q_T$.  
The tree level expressions for $A(Q_T)$ and ${\cal A}(Q_T)$ will be denoted by
$A^{(0)}(Q_T)$ and ${\cal A}^{(0)}(Q_T)$. They are given by 
(ignoring electroweak interference effects for simplicity)
\beq
A^{(0)}(Q_T^{}) = \frac{Q_T^2 R^2 \exp(-R^2 Q_T^2/2)}{4 M^2 R_u^2 \exp(-R_u^2
Q_T^2/2)}\; \frac{\sin^2 \theta_2}{1 +
\cos^2\theta_2}\; 
\frac{\sum_{a}\;e_a^2\; H_1^{\perp a}(z_1) \; \overline H{}_1^{\perp a}(z_2)}{
\sum_{b}\; e_b^2\; D_1^b(z_1) \; \overline D{}_1^b(z_2)}, 
\label{treelevelA}
\eeq 
and  
\beq 
{\cal A}^{(0)}(Q_T) = 
\exp\left[-(R^2-R_u^2)Q_T^2/2\right] M^2 Q_T^2 R^6/R_u^2.
\eeq 
\begin{figure}[htb]
\begin{center}
\leavevmode \epsfxsize=10cm \epsfbox{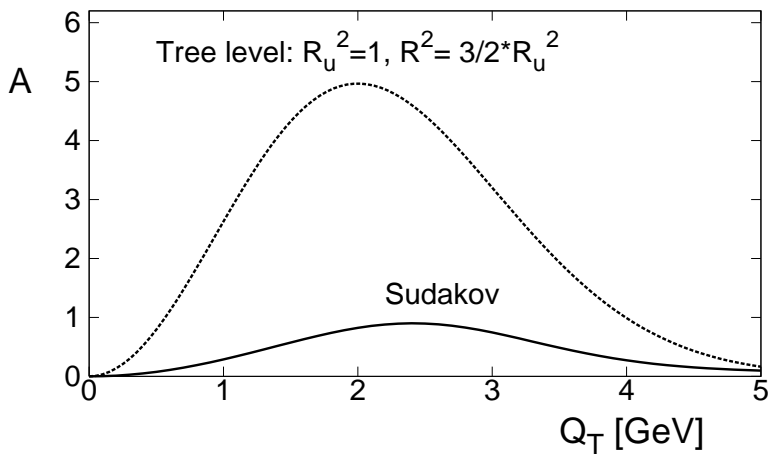}
\caption{\label{4Belle2}The asymmetry factor ${\cal A}(Q_T)$ at 
$Q = 10 \, \text{GeV}$ (solid curve) 
and the tree level quantity ${\cal A}^{(0)}(Q_T)$ 
using $R_u^2=1 \, \text{GeV}^{-2}$ and
$R^2/R_u^2 = 3/2$. Both factors are given in units of
$M^2$.}  
\end{center}
\vspace{-2 mm}
\end{figure}
In Fig.\ \ref{4Belle2} we have displayed the comparison of ${\cal A}(Q_T)$ at
$Q = 10 \, \text{GeV}$ and the tree level
quantity ${\cal A}^{(0)}(Q_T)$ using the values $R_u^2=1 \, \text{GeV}^{-2}$
and $R^2/R_u^2 = 3/2$, which were chosen such as to minimize the magnitude of
${\cal A}^{(0)}(Q_T)$, cf.\ \cite{Boer01} for further discussion.  We conclude
that inclusion of Sudakov factors has the effect of suppressing the tree level
result roughly by a factor of 5, 
whereas for $Q=90 \,
\text{GeV}$ it is more than an order of magnitude. Tree level extractions of
the Collins function at
large $Q^2$ therefore can significantly underestimate its actual magnitude
(roughly by the square-root of the Sudakov suppression factor of the
asymmetry). It is important to keep
this in mind when comparing predictions or fits of transverse momentum
dependent azimuthal spin asymmetries based on tree level expressions applied
at different energies. 

The above also shows that upon including Sudakov
factors one retrieves parton model or tree level 
characteristics (also noted in Ref.\
\cite{CS85e}), but with transverse momentum spreads that are significantly
larger than would be expected from intrinsic transverse momentum (this is
supported by the presently available parameterizations of $S_{NP}$ in various
processes, which usually have Gaussian $b$-dependence with widths that
increase with $Q^2$, cf.\ e.g.\ \cite{Meng-95}). 

\subsection{\label{sec:SNP}Nonperturbative Sudakov factor from BELLE}

Since the previous results depend on the input for the nonperturbative Sudakov
factors $S_{NP}$, which (as a function of $z_1, z_2$) is not determined for the
process $e^+ e^- \to h_1\; h_2 \; X$, the numerical conclusions about the size
and $Q^2$ dependence of the suppression should be viewed as generic, not as
precise predictions. Therefore, we would like to stress the need for an
extraction of the nonperturbative Sudakov factor from the process $e^+ e^- \to
h_1\; h_2 \; X$.  Considering the wealth of data from BELLE this should pose
no problem. Here we will give a brief outline of how this could be done.

Thus far the nonperturbative Sudakov factor has only been obtained from older
$e^+ e^- \to h_1\; h_2 \; X$ data for the energy-energy correlation 
function \cite{Basham:1978zq}, 
$\frac{1}{8}\sum_{h_1,h_2} \int dz_1 z_1 dz_2 z_2 Q^2 
d\sigma/dz_1dz_2dQ_T^2$, at various values of $Q^2$ \cite{CS-PRL82}. A method
based on CSS factorization \cite{CSS-85} was used as discussed in detail in 
Ref.\ \cite{CS85e}. CSS factorization can be obtained from CS factorization 
for the $\phi$-integrated cross section which receives only contributions from
unpolarized quarks (cf.\ the discussion in Ref.\ \cite{Bacchetta:2008xw} 
on the relation between CSS and CS factorization for SIDIS). 
In CSS factorization only collinear 
fragmentation functions appear, which do not affect the energy-energy 
correlation 
function on account of the momentum sum rule $\sum_h \int dz z D_1^{a \to
  h}(z)=1$.   

It needs to be mentioned however that the $Q_T$ distribution for the Drell-Yan 
process requires an $S_{NP}$ that is dependent on the momentum fractions 
\cite{Landry}. It would therefore be better to extract $S_{NP}$ from 
$e^+ e^- \to h_1\; h_2 \; X$ data at given $z_1$ and $z_2$, rather than from the
energy-energy correlation function. 

To obtain the nonperturbative Sudakov factor the cross section differential 
in $z_1$, $z_2$ and $Q_T$ needs to be fitted (as said, one can integrate over
$\phi_1$). Either one uses the CSS formalism as explained in Ref.\
\cite{CS85e} or one can use the CS formalism with a suitable $b$
dependence of the fragmentation functions. Here we will do the latter and 
stay within the pragmatic approach adopted before, employing a Gaussian form 
(\ref{GaussFT}). In this way we arrive at 
(leaving the $Q_0$ dependence implicit) 
\beq
\frac{d\sigma (e^+e^-\to h_1h_2X)}{dz_1 dz_2 dQ_T^{2}} =  
\frac{\alpha^2}{Q^2} \, \sum_{a,\bar{a}} e_a^2 D_1^a(z_1)
\overline{D}{}^a_1(z_2) \, \int_0^\infty db \, b \, J_0(b Q_T)\, 
\exp\left({-S(b_*,Q)}\, {-S_{NP}(b,Q)} \right) \, \tilde{U}(b_*).
\eeq
This expression together with Eq.\ (\ref{sudactual}) for $S(b)$ 
and the general parameterization of $S_{NP}$ in Eq.\ (\ref{genNPform}) 
can be used to obtain a fit of the parameters $g_i$ to the data at 
small $Q_T$ ($\ll
Q^2$, in order to avoid inclusion of ${\cal O}(Q_T^2/Q^2)$ terms 
that were dropped in Eq.\ (\ref{CS81xs})). Such an extraction would be of 
more general interest as well,
since there is a relation between the dominant $Q^2$-dependent part of
nonperturbative Sudakov
factors in the Drell-Yan process, SIDIS and $e^+ e^- \to h_1 \; h_2\; X$
\cite{Meng-95}, which would be interesting to test.

\section{\label{sec:Rad}Radiative corrections}

Thus far we have restricted the discussion to $Q_T^2 \ll Q^2$. In this section
we will look at the high-$Q_T$ region, where one expects collinear
factorization and fixed order perturbation theory to yield a good description
of the cross section.

\begin{figure}[htb]
\begin{center}
\leavevmode \epsfxsize=10cm \epsfbox{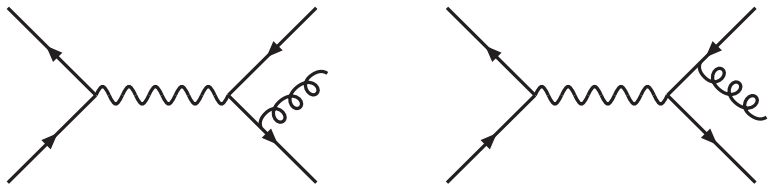}
\vspace{0.2 cm}
\caption{\label{fig:Belle} Lowest order gluon bremsstrahlung diagrams.}
\vspace{-5 mm}
\end{center}
\end{figure}

It is well known that gluon bremsstrahlung also leads to angular asymmetries,
as the radiating quark or antiquark is recoiling away from the original
quark-antiquark axis. In this section we will look at this effect at order
$\alpha_s$, i.e.\ the radiation of an additional gluon into the final state:
$e^+ \; e^- \to q \; \bar{q} \; g$ (we will not include $\gamma$-$Z$
interference terms; this can be done at a later stage, when required), see
Fig.\ \ref{fig:Belle}.  As it turns out the expressions are the simplest in
the Collins-Soper frame. This frame is the lepton-pair center of mass frame
with the $\hat{\bm{z}}$ axis defined as pointing in the direction that bisects
the three-vectors $\bm{P}_2$ and $-\bm{P}_1$ and was discussed in Sec.\
\ref{sec:Frames}.

In the collinear factorization approach contributions to $W_T, W_L, W_\Delta$
and $W_{\Delta \Delta}$ will be generated at order $\alpha_s$. In the CS
frame certain ratios of these structure functions will be independent of the
subsequent fragmentation of the quark and antiquark. The expressions we will
give here are completely analogous to the Drell-Yan expressions obtained in
Ref.\ \cite{Collins:1978yt}. They are valid for large $Q_T$ ($\gg M$), but
still restricting to two jet events. Here we will not consider the case where
one of the hadrons comes from the gluon fragmenting, which becomes relevant
when $Q_T \sim Q$. For pronounced three jet events it would certainly have to
be taken into account.

At large $Q_T$ at least one of the hadron momenta is deviating considerably
from the $\hat{\bm{z}}$ direction. The hadron momenta are again $P_1$ and $P_2$.
Upon neglecting $M_i^2/Q^2$ corrections, the photon momentum is 
parameterized as $q= P_1/\zeta_1 + P_2/\zeta_2 + q_T$ (cf.\ Eqs.\
(\ref{P1})-(\ref{q})), 
such that $Q^2 = \tilde{s}/(\zeta_1\zeta_2) - Q_T^2$, where 
$\tilde{s}= (P_1+P_2)^2$. It follows that $z_i= \zeta_i \; \tilde{Q}^2/Q^2$.  
The calculation is done in
collinear factorization, which means that parton transverse momenta w.r.t.\
the produced hadrons are neglected. The parton momenta are defined such that
$p_i = P_i/\xi_i$.

Following the calculation of \cite{Collins:1978yt}, but now applied to process 
$e^+ \; e^- \to q \; \bar{q} \; g \to h_1 \; h_2 \; X$ 
(that is, one has to calculate the contribution of the 
two diagrams in Fig.\ \ref{fig:Belle}), 
one finds {\em in the CS frame\/} 
\ba
\lefteqn{\frac{dN}{d\Omega}  =  
\frac{3}{16\pi} \left[ \frac{Q^2 + \frac{3}{2} Q_T^2}{Q^2+Q_T^2} 
+ \frac{Q^2 - \frac{1}{2} Q_T^2}{Q^2+Q_T^2} \cos^2\theta_{CS}
\right.}\nn\\[2 mm]
&& \quad \left. \mbox{} 
+ \frac{Q_T Q}{Q^2+Q_T^2} K(\zeta_1,\zeta_2,Q_T^2/\tilde{s})
\sin 2\theta_{CS} \cos \phi_{CS} 
+ \frac{1}{2}\frac{Q_T^2}{Q^2+Q_T^2} \sin^2 \theta_{CS}
\cos 2\phi_{CS} \right],
\label{Collins79}
\ea
where the function $K(\zeta_1,\zeta_2,Q_T^2/\tilde{s})$ is given by
\beq
K(\zeta_1,\zeta_2,Q_T^2/\tilde{s}) = 
\frac{\int d\xi_1\int d\xi_2\; \delta\left(
  (1/\xi_1-1/\zeta_1)(1/\xi_2-1/\zeta_2) - Q_T^2/\tilde{s}\right) 
\sum_a e_a^2 D_1^a(\xi_1,\mu)
  \overline D{}_1^{a}(\xi_2,\mu)\left(\zeta_2^2/\xi_2^2 -\zeta_1^2/\xi_1^2
\right)}{\int d\xi_1\int d\xi_2\; \delta\left(
  (1/\xi_1-1/\zeta_1)(1/\xi_2-1/\zeta_2) - Q_T^2/\tilde{s}\right) 
\sum_a e_a^2 D_1^a(\xi_1,\mu)
  \overline D{}_1^{a}(\xi_2,\mu)\left(\zeta_2^2/\xi_2^2 +\zeta_1^2/\xi_1^2
\right)} \ .
\eeq 
This expression shows that only the partonic
$\hat{W}_\Delta$ depends on the lightcone momentum fractions of partons and
therefore that only the hadronic $W_\Delta$ will depend on the fragmentation
functions $D_1$ and $\overline D_1$.  This is only the case in the
Collins-Soper frame. In the Gottfried-Jackson frame all angular dependences
will depend on the fragmentation functions. 

Using the other standard notation for the angular dependences 
Eq.\ (\ref{lmnnotation}) one finds that the above yields:
\ba
\lambda_{CS} & = & \frac{Q^2-\frac{1}{2}Q_T^2}{Q^2+\frac{3}{2}Q_T^2}\; , 
\\
\nu_{CS} & = & \frac{Q_T^2}{Q^2+\frac{3}{2}Q_T^2} \ .
\label{LOform}
\ea
We see that also in $e^+ e^- \to h_1 \; h_2 \; X$ the so-called Lam-Tung 
relation holds~\cite{Lam-78,Lam-80}: 
\beq
1-\lambda_{CS}-2\nu_{CS}=0 \ ,
\eeq
or equivalently, $W_L = 2 W_{\Delta \Delta}$. Using Eq.\ (\ref{CStoGJ}) one
sees that the relation also holds in the GJ frame and as shown for the
Drell-Yan 
process, it is expected that it continues to hold if one includes the gluon 
fragmentation
contribution, although in that case both $\lambda$ and $\nu$ {\em will\/} 
depend on
the fragmentation functions (cf.\ Ref.\ \cite{Boer:2006eq} for a
discussion of these aspects for the Drell-Yan process, including the effects
of resummation). We also note that the Lam-Tung relation is purely
a ${\cal O}(\alpha_s)$ result, i.e.\ it does not apply beyond leading order.
Surprisingly, in the Drell-Yan process the Lam-Tung relation is known to be
violated by much more than the ${\cal O}(\alpha_s^2)$ contribution 
\cite{Brandenburg-93}. The
distribution function analogue of the Collins effect has been suggested as an
explanation for this large deviation from the NLO pQCD result \cite{Boer-99}. 
It would be very interesting to check experimentally in
$e^+ e^- \to h_1 \; h_2 \; X$ whether the Lam-Tung relation is
violated as much as in the Drell-Yan process. 

A natural question is how to combine this collinear factorization
fixed order perturbative result, which is valid for $M^2 \ll Q_T^2$,
with the CS factorization Collins effect result, which is valid for
$Q_T^2 \ll Q^2$. This question was addressed recently in Ref.\ 
\cite{Bacchetta:2008xw} for the analogous situation in semi-inclusive
DIS. There it was pointed out that the two results may simply be
added, which gives an expression for the asymmetry that is correct up
to power suppressed corrections (of order $Q_T^2/Q^2$ in the low $Q_T$
region and of order $M^2/Q_T^2$ in the high $Q_T$ region). It is based
on the fact that the denominators of the asymmetry expressions in Eqs.\
(\ref{nufromCollins}) and (\ref{LOform}) coincide in the intermediate
$Q_T$ region ($M^2 \ll Q_T^2 \ll Q^2$) (a fact that is also used in
the derivation of the CSS formalism \cite{CSS-85} from CS
factorization). 
In addition, one uses the following observations. The above calculation
shows that the fixed order perturbative expression for $\nu$ is of
order $Q_T^2/Q^2$ when $Q_T^2 \ll Q^2$, hence power suppressed w.r.t.\ 
the Collins asymmetry expression given in Eq.\ (\ref{nufromCollins}).
Conversely, considering the latter in the large $Q_T$ ($\gg M$) limit
beyond tree level (within CS factorization) leads to a result of order
$M^2/Q_T^2$ \cite{Bacchetta:2008xw}, 
which is power suppressed w.r.t.\ $\nu$ in Eq.\ 
(\ref{LOform}). This implies that upon neglecting power suppressed
terms the sum of the two contributions reduces to Eq.\ 
(\ref{nufromCollins}) at low $Q_T$ and to Eq.\ (\ref{LOform}) at high
$Q_T$. In the intermediate region the results can be added because
they have a common denominator. For further details we refer to 
\cite{Bacchetta:2008xw}.

Despite the fact that the asymmetry for all $Q_T$ can be described by
the sum of the Collins effect expression and the fixed order collinear
factorization expression, the latter complicates the extraction of
the Collins function and the weighting with $Q_T^2$, as explained
next. Therefore, one wants to construct quantities 
in which the hard gluon radiation component
of the $\cos 2\phi$ asymmetry is (largely) absent. One such option is
to consider 
ratios of asymmetries where the hard gluon radiation component
of the $\cos 2\phi$ asymmetry drops out. The other would be to exploit the
Lam-Tung relation. Both options will be discussed below. 

\section{\label{sec:Weigh2}Weighted asymmetry beyond tree level}

In this section we return to the issue of integrating over $Q_T$ and weighting
with $Q_T^2$. In the cross section kept differential in the transverse
momentum $\bm{q}_T^{}$ the Collins effect gives rise to a $\cos 2 \phi_1$
asymmetry (both in the GJ and CS frames). In principle one can consider this
quantity integrated over the length $Q_T$ of $\bm{q}_T^{}$, but this yields a
quantity that does not arise in other processes in the same way. Without
further assumptions about the Collins fragmentation function, such a quantity
would only be useful to test whether it has the expected $Q^2$ behavior. This
formed the main reason for the suggestion to weight the $Q_T^2$ integration
with an additional factor of $Q_T^2$. At tree level this yields an expression
involving the first transverse moment of the Collins function.  Beyond tree
level but considering only the low-$Q_T$ expression of the CS formalism
this remains true and moreover, a quantity results that does not suffer from
Sudakov suppression\footnote{One probes the Sudakov factor at the point $b=0$
  where it vanishes, although that cannot be seen from the perturbative
  expression for $S(b)$ in Eq.\ (\ref{sud}), which is only valid for $b >
  b_0/Q$.}, as was pointed out in Ref.\ \cite{Boer01}. These nice properties
are spoiled however when the perturbative high-$Q_T$ contribution is included.
Unfortunately the weighted cross section becomes sensitive mainly
to the large $Q_T^{}$ contribution (note that we are discussing the asymmetry
in the weighted cross section, i.e.\ we integrate numerator and denominator of
the $Q_T$-dependent asymmetry separately).
Because the latter is calculable perturbatively and can
simply be added to the low-$Q_T$ result, it can in principle be
subtracted from the experimental result before extracting $H_1^{\perp (1)}$. 

One way to subtract it experimentally is to consider 
ratios of asymmetries where the hard gluon radiation component
of the $\cos 2\phi$ asymmetry drops out. This is discussed in the next
section. The other possibility would be to exploit the
Lam-Tung relation. Here the idea is to
use the structure functions themselves, as opposed to the ratios of
structure functions $\lambda$ and
$\nu$. We observe that the quantity
\beq 
\int d^2 \bm{q}_T^{} \bm{q}_T^2 \left(W_L -2 W_{\Delta \Delta} \right) 
= \frac{8M_1M_2}{z_1^2 z_2^2} \sum_{a,\bar a}e_a^2\; 
\;H_1^{\perp(1)a}(z_1) \,\overline H_1^{\perp (1)a}(z_2)
\eeq
does not receive radiative corrections at order $\alpha_s$. This
applies in both the GJ and CS frames and will hold true even upon
inclusion of the gluon fragmentation contribution.  

Once the first transverse moment of the Collins function has been extracted 
there is another potentially interesting test of the
Collins effect in the CS formalism, which may turn out to be scale 
independent. This exploits the 
so-called Sch\"afer-Teryaev sum rule \cite{Schafer:1999kn} 
\beq
\sum_h \int dz\;  z \; H_1^{\perp (1)}(z) = 0.
\eeq
If we define for each hadron type $h$ the function $C_h$ that is a
function of $M_h$ and scale $\mu$ only, as
\beq
C_h(M_h,\mu) \equiv \int dz\;  z \; H_1^{\perp (1)}(z),
\eeq
then one can conclude that for each hadron type this has the same
$\mu$ dependence (if any). This is because there is no gluon 
contribution, such that there is no mixing among gluon and quark
functions. For each quark flavor one has the same evolution
properties. Assuming autonomous evolution of the function 
$H_1^{\perp (1)}(z)$, this implies that ratios of $C_h$'s for 
different hadrons types, 
e.g.\ $C_\pi/C_K$, will be scale independent. This could then 
be tested experimentally without the need to know the scale dependence of the
Collins asymmetries. Measuring the ratios at BELLE, HERMES, LEP1,
etc, should then all give the same answer. The only experimental 
difficulty would be to obtain
a sufficient coverage in $z$ and $Q_T$ to be able to extract the
integrated quantities $C_h$ reliably. 
In Ref.\ \cite{Henneman:2001ev} 
an autonomous evolution equation for $z H_1^{\perp (1)}(z)$ has been obtained
in the large $N_c$ limit. However, since its derivation employed 
so-called Lorentz invariance relations the correctness of which has
afterwards been disputed \cite{Goeke:2003az}, 
the assumption of autonomous evolution may turn out to be unwarranted. This
issue should first be settled before any conclusions can be drawn from the
ratios $C_{h_1}/C_{h_2}$. But if scale independence of the ratios 
$C_{h_1}/C_{h_2}$ can be established, it would offer 
an experimental consistency check of the extracted Collins fragmentation 
functions. 

\section{\label{sec:Ratios}Ratios of asymmetries}

There are several reasons why it may be useful to consider taking ratios of
asymmetries for different sets of hadrons. These may be experimental reasons
such as to cancel systematic effects, but also theoretical reasons
namely that effects drop out that are independent of the type of hadron
considered.

If the $\cos 2\phi$ asymmetry itself is not that large, the asymmetry 
in a ratio of two $\cos 2\phi$ asymmetries is to good
approximation also a $\cos 2 \phi$ asymmetry:
\beq
\frac{\frac{d\sigma (e^+e^-\to h_1\; h_2\; X)}{dz_1 dz_2 d\Omega d^2{\bm
      q_T^{}}}}{\frac{d\sigma (e^+e^-\to h_3\; h_4\; X)}{dz_1
    dz_2 d\Omega d^2{\bm q_T^{}}}} \propto \frac{ 
1 + \cos 2\phi_1 A^{h_1 h_2}(\bm{q}_T^{}) }{ 
1 + \cos 2\phi_1 A^{h_3 h_4}(\bm{q}_T^{}) } \approx 
1 + \cos 2\phi_1 \left\{A^{h_1 h_2}(\bm{q}_T^{}) - A^{h_3
    h_4}(\bm{q}_T^{})\right\}.
\label{doubleratio}
\eeq
As can be seen, effects that are independent of the type of
hadrons $h_i$ cancel in the difference $A^{h_1 h_2} - A^{h_3 h_4}$. 
Since fragmentation functions generally do depend on the type of hadron, only
effects that are independent of the fragmentation functions will
cancel. In Sec.\ \ref{sec:Rad} 
it was discussed that in the Collins-Soper frame the
perturbative order-$\alpha_s$ contribution to $\nu$ (Eq.\ (\ref{LOform})) is
independent of the fragmentation functions. This assumes that the hadrons
arise from quark fragmentation, which should be a good approximation for $Q_T$
not too close to $Q$. Hence, under these conditions (in the CS frame for
moderate $Q_T$) the order-$\alpha_s$ radiative correction term cancels in the 
ratio Eq.\ (\ref{doubleratio}). If in addition $\mu_{CS} \approx 0$, then the 
conclusion
also extends to other frames that are rotations of the CS frame, such as the
GJ frame or the jet frame discussed in Sec.\ \ref{sec:Jet}.  

Since at moderate $Q_T$ the dominant contribution to $\nu$ from radiative
corrections (Eq.\ (\ref{LOform})) is already small itself, any uncanceled
remainder from higher order corrections or from the gluon fragmentation
contribution is expected to be negligible. This can be verified via explicit
prediction of the radiative correction to the cross section, e.g.\ via a
Monte-Carlo study. Due to the cancellation of the dominant radiative
correction, the ratio method presents a good way to access the Collins
effect without taking into account the radiative corrections explicitly. The
use of different types of hadron pairs, such as $\pi^+ \pi^-$, $\pi^\pm
\pi^\pm$ and $\pi^{\pm} \pi^0$, moreover allows one to learn about favored
versus unfavored fragmentation functions \cite{Efremov:2006qm}.

Predictions for the ratio of jet frame 
$\cos(\phi + \phi')$ asymmetries (cf.\ Eq.\
(\ref{phiphiprime}) or Ref.\ \cite{Anselmino:2007fs}) for unlike-sign pion
pairs to like-sign pairs have been given in Ref.\ \cite{Bacchetta:2007wc}
employing a spectator model. Asymmetry ratio's of the order of the
experimental data could be obtained, but with considerable
uncertainties. Interestingly, kaon
asymmetries were predicted to be larger than the pion asymmetries.
 
\section{\label{sec:Beampol}Beam polarization}

Thus far we have assumed that no transverse beam polarization is present. 
However, it is well-known that charged particles circulating in a 
magnetic field become polarized transversely to the beam direction due to
emission of spin-flipping synchrotron radiation: 
the Sokolov-Ternov effect \cite{Sokolov:1963zn}. 
This effect can be significant for electrons and
positrons due to their small mass. The transverse polarization of particles
circulating in a uniform magnetic field is the following function of time:
\beq
P(t) = \frac{1}{a} 
\left\{1- \exp\left(-a \frac{e^2 \hbar \gamma^5}{m^2 c^2 \rho^3}
    t\right) \right\},
\label{Sokolov}
\eeq
where $\gamma=E/m$ is the Lorentz factor, $\rho$ is the bending radius, and 
$a=5\sqrt{3}/8$. More general situations are reviewed in Ref.\ 
\cite{Jackson:1975qi}. The polarization has a strong dependence on the mass in 
the exponent: the lighter the particle, the faster it becomes polarized.  
 
It is also well-known \cite{Schwitters,Close,Schiller:1979cv,Olsen:1980cw}
that beam polarization can affect the angular distribution of produced hadrons
in $e^+ e^- \to h X$. As explained in \cite{Close} it leads to a $\cos 2\phi$
asymmetry, where the angle $\phi$ is defined w.r.t.\ the beam and beam spin
directions. This asymmetry has in common with the Collins effect asymmetry
that both are transverse spin asymmetries: the former concerns lepton spins,
the latter quark spins. Both asymmetries arise due to interference between
contributions of $\pm 1$ photon helicity states. In the former the magnetic
field determines the electron and positron spin directions, which then
determine the photon polarization state and hence, the subsequent decay of the
photon into quark-antiquark pairs, which is visible in the angular
distribution of final state hadrons.  In the Collins asymmetry case, the
angular distribution of two final state hadrons w.r.t.\ each other filters or
selects the quark and anti-quark spin directions, which via the photon are
correlated to the electron-positron angular distribution. The Collins effect
does not contribute to the angular distribution of produced hadrons in 
$e^+ e^- \to h X$, but transverse beam polarization may affect the 
angular distribution in $e^+ e^- \to h_1 \; h_2 \; X$.  

Turning to the situation at BELLE: one should learn whether the beams
are polarized due to the Sokolov-Ternov effect, because it may lead to
angular asymmetries that require correcting for. From the beam energies and
bending radii of the KEK accelerator one can try to estimate the polarization
build-up time from Eq.\ (\ref{Sokolov}). Here one should keep in mind that
the accelerator consists mostly of straight sections. For the low (high)
energy ring the bending radius is 16.3 m (104.5 m), but the length of bending
is 0.915 m (5.86 m), whereas the circumference of the rings is a little over
3000 m. Therefore, the polarization build-up time interval is much reduced 
compared to simply using the quoted 
bending radii in Eq.\ (\ref{Sokolov}). Moreover, 
depolarization effects should be very important, especially since the
interaction point is in the middle of a straight section. Negligible beam
polarization at the collision point is thus expected. 

The degree of polarization $P$ can be measured experimentally 
via the process $e^+ e^- \to \mu^+ \mu^-$ for which the cross
section is \cite{Close}
\beq
\frac{d\sigma(e^+ e^- \to \mu^+ \mu^-)}{d\Omega} = \frac{1}{2} \sigma_T \left(
1+\cos^2 \theta + P^2 \sin^2\theta \cos 2\phi  \right),
\eeq
where $\sigma_L/\sigma_T = 4m_\mu^2/Q^2 \approx 0$ is used. 
Experimental investigations of this cross section can show the presence or 
absence of any significant transverse beam polarization at BELLE.

\section{\label{sec:Weak2}Weak decays background}

Another test of systematic effects is offered by the process
$e^+ e^- \to \tau^+ \tau^- \to \pi^+ \bar \nu_\tau \pi^- \nu_\tau$ (and
other such weak decays). In this
process a $\cos 2\phi$ asymmetry analogous to the Collins effect asymmetry 
can arise. In this case it is calculable within the Standard
Model, using the parity violation parameters (usually denoted by $\alpha$) 
for the tau decays ($\alpha_{\pi^\pm}=\mp 1$ for $\tau^\pm \to \pi^\pm
\stackrel{(-)}{\nu}_\tau$). The self-analyzing parity-violating weak 
decay of the $\tau$'s plays a role similar to the Collins effect. It
correlates the spatial distribution of the produced pions to the 
spins of the $\tau$'s. The asymmetry does not depend on the 
$\tau$'s being transversely polarized {\em on average\/} (in that case the
effect would have been of order $m_{\tau}^2/s$). 

The analyzing power of the $\cos 2\phi$ asymmetry is not a very simple
expression, because it depends on the decay kinematics, for details we refer
to Refs.\ \cite{Bernabeu:1989ct,Bernabeu:1990na,Alemany:1991ki}. 
The asymmetry is proportional to $\alpha_{\pi^+} \alpha_{\pi^-} C_{TT}
= - C_{TT}$, where 
$C_{TT}$
is given in terms of the axial and vector coupling of the $\tau$ lepton
to the neutral gauge boson: 
\beq
C_{TT} = \frac{|a_\tau|^2-|v_\tau|^2}{|a_\tau|^2+|v_\tau|^2}, 
\label{CTT}
\eeq
which for the $Z$ boson is close to $+1$ and for the photon is $-1$. This
change of sign implies a sign difference between the asymmetry measured at LEP 
\cite{Abreu:1997vp,Barate:1997mz,GarciaAbia:1997xg} and what would be 
obtained at BELLE. The LEP results are in good 
agreement with the standard model
prediction, therefore, assuming the same applies to BELLE, it could 
provide a way to check for systematic effects. 
The same would apply to heavy-quark weak decays.   

\section{\label{sec:Sum}Summary}

We have given an overview of the Collins effect asymmetry in the
process $e^+e^- \to h_1
\, h_2\, X$, where the two final state hadrons belong to opposite jets. We
restricted the discussion to two jet events and discussed the particular
situation of BELLE, but many results also are applicable at other $e^+ e^-$
collider facilities. The Collins effect --a spin effect in the quark 
fragmentation process-- gives rise to an azimuthal asymmetry, which in the
Gottfried-Jackson or Collins-Soper frames is a $\cos 2 \phi$ asymmetry. 
It was our main objective to study to which extent the analyzing
power of this angular asymmetry is a measure of the Collins effect. 

At low values of the observed transverse momentum $Q_T$ of the photon we
employ Collins-Soper factorization, which leads to expressions involving
parton transverse momentum dependent fragmentation functions. It is within
this framework that the Collins effect fragmentation function arises. 
At tree level and leading twist it leads to the only contribution to the 
$\cos 2 \phi$ asymmetry. Twist-3
effects (order $M/Q$ terms) were shown in the Gottfried-Jackson frame to only
contribute to $\mu$, i.e.\ to give rise to a $\cos \phi$ asymmetry.
Throughout this paper we have neglected twist-4 effects that are of order
$M^2/Q^2$. They can contribute to the $\cos 2 \phi$ asymmetry, but are 
expected to lead to $\nu < \mu$, i.e.\ a larger $\cos \phi$ than $\cos 2
  \phi$ asymmetry. This property allows to estimate their importance at
BELLE.

Beyond tree level, but still at low $Q_T$, the Collins-Soper factorization
expression also dictates the scale $Q^2$ dependence of the Collins effect
asymmetry. This was studied extensively and it was demonstrated that the
Collins effect asymmetry suffers from considerable Sudakov suppression as
$Q^2$ increases. This was already pointed out in Ref.\ \cite{Boer01}, but the
analysis presented here improves on that treatment and leads to quantitatively
different but qualitatively very similar results. We conclude that inclusion
of the dominant double and single logarithmic terms at order $\alpha_s$ has
the effect of suppressing the tree level result. In the example considered
this suppression was roughly by a factor of 5 for $Q=10$ GeV as relevant for
the BELLE experiment. Therefore, tree level extractions of the Collins
function can significantly underestimate its actual magnitude. It is important
to keep this in mind when comparing predictions or fits of transverse momentum
dependent azimuthal spin asymmetries based on tree level expressions but
applied at different energies. To properly describe the cross section beyond
tree level within the Collins-Soper formalism, a good experimental
determination of the nonperturbative Sudakov factor is required. An
explanation of how this can be done at BELLE was given.

In the Collins-Soper formalism the Collins fragmentation functions enter in a
transverse momentum convolution expression. Such convolutions do not arise in
other processes in the same way, therefore one has to resort to other means of
extracting information that can be applied in other processes. Often a
Gaussian transverse momentum dependence of the Collins function is assumed
because in that case all integrations can be performed analytically. But this
introduces a model dependence in the results. Another way suggested is to
consider weighted integration over the observed transverse momentum $Q_T$. In
this way the $Q_T^2$-weighted cross section {\em at tree level\/} probes the
first transverse moment of the Collins fragmentation function, which also
arises in other processes identically. This would allow for model independent
predictions once the Collins functions are extracted from the BELLE data.
Another advantage of the $Q_T^2$-weighted cross section is that it does not
suffer from the Sudakov suppression mentioned above. However, considering
the integral over all $Q_T$ requires that one also describes the asymmetry
well at high $Q_T$. This complicates matters considerably.

For the $\cos 2 \phi$ asymmetry at high $Q_T$ collinear factorization can be
used to describe it. This is most conveniently done in the Collins-Soper
frame, where the dependence on the fragmentation functions drops out to a
large extent. The asymmetry arises from hard gluon emission and is not related
to the Collins effect.  Therefore, it becomes important to describe this
dependence, that behaves as $Q_T^2/Q^2$, as best as possible in order to
distinguish it from the Collins effect contribution that at large $Q_T$
behaves as $M^2/Q_T^4$ (as opposed to Gaussian transverse momentum).  Due to
this specific $Q_T^2$ dependence of the two effects, it is possible (upon
ignoring twist-4 contributions) to simply add the two contributions (cf.\
Ref.\ \cite{Bacchetta:2008xw}). The Collins effect then dominates at low $Q_T$
and the hard gluon emission at large $Q_T$. Unfortunately this implies that
the $Q_T^2$-weighted integration is sensitive mostly to the latter effect and
not to the Collins effect of interest. So one first has to subtract the
perturbative collinear factorization result at large $Q_T$. This can be done
to a large extent automatically by considering ratios of asymmetries, but a
new way exploiting the analogue of the Lam-Tung relation was pointed out.
This relation was shown to be violated strongly in the Drell-Yan process
w.r.t.\ the NLO pQCD result. It would be very interesting to see whether this
also holds in $e^+ e^-$ collisions.

To get an idea about the size of the Collins effect asymmetry in $e^+e^- \to
h_1 \, h_2\, X$, a crude estimate of the $Q_T^2$-weighted Collins asymmetry at
tree level was made. It shows the asymmetry can be on the level of a few
percent. This is in agreement with a model prediction in the literature. These
estimates employ tree level expressions, but they should remain reasonable
estimates beyond tree level (due to the absence of Sudakov suppression),
provided the dominant high-$Q_T$ gluon emission contribution is subtracted.

A comparison to the Collins asymmetry in the so-called jet frame (defined with
help of the $q \bar{q}$ axis or approximately by the thrust axis) was given.
In this frame the Collins asymmetry becomes a $\cos (\phi+\phi')$ asymmetry.
It was shown to probe so-called half-moments of the Collins fragmentation
functions, as opposed to the first transverse moments appearing in the
$Q_T^2$-weighted Collins asymmetry. It is estimated to be similar in size as
the asymmetry in the GJ frame. 

Other potential contributions to the Collins effect asymmetry 
have been investigated, such as from $\gamma$-$Z$ interference and the effect
of beam polarization. Both effects should be negligible at BELLE (well below
the percent level). Electroweak contributions can in principle 
lead to a $\sin 2\phi_1$ asymmetry, but also that was estimated to be very
small. Therefore, we conclude that $\gamma$-$Z$ interference does not lead to 
any significant additional $\phi$ dependence. It only modifies the $\theta$ 
distribution with a forward-backward asymmetry term of a few percent.  

A $\cos 2\phi$ asymmetry analogous to the Collins effect asymmetry arises in 
$e^+ e^- \to \tau^+ \tau^- \to \pi^+ \bar \nu_\tau \pi^- \nu_\tau$ (and
other such weak decays). Using the known $\tau$ weak decay parameters 
this asymmetry can be calculated entirely within the electroweak sector of
the Standard Model. A comparison of the data to this Standard Model
result could provide a good way to 
check for systematic effects. 

Based on all these considerations it appears feasible to arrive at an actual  
measurement of the Collins effect and an extraction of the Collins
fragmentation function to a reasonable extent. 
Based on arguments in favor of the universality of the Collins fragmentation 
function for the
processes $e^+ e^- \to h_1 h_2 X$ and SIDIS, a simultaneous fit to the Collins
effect asymmetry data can be and already has been performed, leading to a first
extraction of transversity \cite{Anselmino:2007fs}. This 
demonstrates clearly and explicitly the merit of the Collins effect
asymmetry measurement at BELLE. 

\begin{acknowledgments}
I would like to thank Rainer Jakob and Piet Mulders for past collaboration on
this topic. Also, I am grateful to Matthias Grosse Perdekamp, Kazumi Hasuko,
Jens Soeren Lange, Akio Ogawa and Ralf Seidl for numerous 
discussions that were essential for this overview to take shape. 
I also acknowledge fruitful comments by Bruce Yabsley. 
Finally, I wish to thank Alessandro Bacchetta, John Collins, Umberto D'Alesio, 
Anatoli Efremov and Werner Vogelsang for discussions on some of 
the theoretical aspects of this study and Markus Diehl for discussions and
extensive feedback on the manuscript.
\end{acknowledgments}


\end{document}